\documentclass[journal=jctcce,manuscript=article]{achemso}

\usepackage{yhmath}
\usepackage{amssymb} 
\usepackage[version=3]{mhchem}
\usepackage[normalem]{ulem}
\usepackage{cancel}
\usepackage{physics}
\usepackage{threeparttable}
\usepackage{multirow}
\usepackage{booktabs}
\usepackage{xcolor}
\usepackage[ruled,vlined,linesnumbered]{algorithm2e}
\SetKw{Continue}{continue}

\newcommand{\ii}{\mathrm{i}}
\newcommand{\ee}{\mathrm{e}}

\newcommand{\QP}{\mathrm{QP}}

\newcommand{\xc}{\mathrm{xc}}

\newcommand{\bvec}[1]{\mathbf{#1}}
\newcommand{\br}{\bvec{r}}
\newcommand{\bq}{\bvec{q}}
\newcommand{\bk}{\bvec{k}}
\newcommand{\bG}{\bvec{G}}

\newcommand{\bx}{\bvec{x}}

\newcommand{\bR}{\bvec{R}}

\newcommand{\Nk}{N_{\bk}}
\newcommand{\invNk}{\frac{1}{\Nk}}
\newcommand{\invsqNk}{\frac{1}{\sqrt{\Nk}}}

\newcommand{\kg}[3]{$#1\times #2 \times #3$}
\newcommand{\kgh}[1]{\kg{#1}{#1}{#1}}

\newcommand{\erfc}{\mathrm{erfc}}

\newcommand{\cMu}{\mathcal{U}} 
\newcommand{\cNu}{\mathcal{V}} 

\newcommand{\Wc}{W^{\mathrm{c}}}
\newcommand{\bWc}{\mathbf{W}^{\mathrm{c}}}

\newcommand\LabelAndRemember[2]
  {\expandafter\gdef\csname labeled:#1\endcsname{#2}%
   \label{#1}#2}
\newcommand\RecallLabel[1]
   {\csname labeled:#1\endcsname\tag{\ref{#1}}}

\newcommand{\JCTC}[1]{#1}

\author{Min-Ye Zhang}
\email{minyez@iphy.ac.cn}
\affiliation{Institute of Physics, Chinese Academy of Sciences, Beijing 100190, China}
\alsoaffiliation{The NOMAD Laboratory at the Fritz Haber Institute of the Max-Planck-Gesellschaft, Berlin 14195, Germany}

\author{Peize Lin}
\affiliation{Institute of Physics, Chinese Academy of Sciences, Beijing 100190, China}
\alsoaffiliation{Institute of Artificial Intelligence, Hefei Comprehensive National Science Center, Hefei 230026, Anhui, China}

\author{Rong Shi}
\affiliation{Institute of Physics, Chinese Academy of Sciences, Beijing 100190, China}

\author{Xinguo Ren}
\email{renxg@iphy.ac.cn}
\affiliation{Institute of Physics, Chinese Academy of Sciences, Beijing 100190, China}

\title[]
  {Low-scaling \textit{GW} calculations of quasi-particle energies for extended systems within the numerical atomic orbital framework}

\abbreviations{MBPT, GW, RI, local approximation}

\begin{document}

\begin{abstract}
The many-body perturbation theory within the $GW$ approximation is a widely used  method for describing the electronic band structures in real materials.
Its application to large-scale systems is, however, impeded by its high computational cost.
The rate-limiting steps in a typical $GW$ implementation are the evaluation of the polarization function under the random phase approximation (RPA) and the evaluation of the $GW$ self-energy, both of which have a canonical $\mathcal{O}(N^4)$ scaling with $N$ being the system size.
The conventional space-time algorithm within the plane-wave basis sets reduces the scaling from $\mathcal{O}(N^4)$ to $\mathcal{O}(N^3)$, albeit with a large prefactor and increased memory cost. Here, we present a space-time algorithm within the numerical atomic orbital (NAO) basis-set framework,
for which the evaluation of the polarization function and self-energy is formally reduced to $\mathcal{O}(N^2)$ or better with respect to system size.
This is achieved by computing these quantities in real space, where low-scaling algorithms can be formulated by leveraging the localized resolution of identity (LRI) technique.
The resulting NAO-based, LRI-enhanced space-time $GW$ algorithm has been implemented in the LibRPA library interfaced with the FHI-aims code package. Benchmark calculations for crystalline solids show that the low-scaling implementation yields quasi-particle energies in close agreement with the conventional $\mathcal{O}(N^4)$ $\bk$-space formalism previously implemented in FHI-aims.
For the systems studied here, the observed overall scaling is substantially reduced relative to the canonical approach,
and the low-scaling implementation becomes advantageous already for systems containing fewer than 100 atoms.
\end{abstract}

\section{Introduction}

First-principles descriptions of electronic band structures are essential for the study of functional materials used in photovoltaic and optoelectronic applications.
Kohn–Sham density functional theory (KS-DFT) \cite{HohenbergP64,KohnW65} offers a widely used approach that balances computational cost and accuracy. However, with common density functional approximations (DFAs), KS-DFT often fails to deliver quantitatively accurate electronic band structures. Conventional local and semi-local DFAs severely underestimate the band gaps of semiconductors and insulators, and may even incorrectly classify narrow-gap semiconductors such as Ge as metals — a limitation commonly known as the band-gap problem \cite{PerdewJ17GKS}. Hybrid functionals formulated within generalized Kohn–Sham theory \cite{SeidlA96GKS} improve the predictive accuracy of band gaps, albeit at the expense of introducing additional adjustable parameters and phenomenological modeling \cite{ZhangMY20WIRES}.
On the other hand, the $GW$ method \cite{HedinL65}, grounded in many-body perturbation theory based on Green's functions, has proven to be a highly accurate approach for describing charged excitations in extended systems and molecules \cite{StrinatiG80,StrinatiG82,HybertsenM85,AryasetiawanF98,OnidaG02-MBPT-rev,GolzeD19GW}.
It has been particularly successful in predicting electronic band structures of weakly and moderately correlated materials, including conventional semiconductors and emerging low-dimensional systems, typically reproducing experimental band gaps within a few hundred meV \cite{KlimesJ14,JiangH16}. Nevertheless, the conventional reciprocal-space $GW$ implementation scales as $\mathcal{O}(N^4)$ with system size and carries a large computational prefactor. Although still more tractable than advanced wave-function-based methods such as the coupled cluster method \cite{MoermanE25,MoermanE25JCTC}, applying $GW$ to systems containing hundreds of atoms or more remains a major challenge.

Much effort has been devoted to circumventing the computational bottleneck in many-body perturbation theory. One prominent approach is the so-called space-time method, introduced by Rojas et al. \cite{RojasH95,RiegerM99}, which avoids convolution in reciprocal space and frequency domain by working with the real-space Green's function in the time domain via fast Fourier transforms, achieving cubic scaling with system size.
To further enhance performance, efficient time and frequency grids based on the minimax quadrature were developed, significantly reducing the number of required grid points.
These grids were first applied to correlation energy calculations within the random-phase approximation,\cite{KaltakM14,KaltakM14Si} and were later extended to $GW$ calculations using plane-wave basis sets \cite{LiuP16}.
The use of localized basis sets represents a promising route to further reduce computational scaling by exploiting matrix sparsity. This direction has been explored with Gaussian-type orbitals \cite{WilhelmJ18,DucheminI21,GramlM24} and Slater-type orbitals.\cite{ForsterA20,ForsterA21}
\JCTC{%
For periodic systems, $GW$ implementations with $\bk$-space sampling based on Gaussian orbitals and the resolution of identity (RI) approximation have also been reported, including both canonical and space-time formulations.\cite{ZhuT21,PasquierR25}
}%
The numerical atomic orbitals (NAOs) provide a more compact and flexible framework of localized basis functions.
Recently, a canonical $GW$ implementation for periodic systems using NAOs \cite{RenX21} was realized in the FHI-aims package \cite{BlumV09} and benchmarked against linearized augmented plane-wave implementations.\cite{JiangH13}
This approach leverages the local variant of the resolution of identity (LRI) technique, which substantially reduces the storage requirements for the RI coefficients. Nevertheless, a low-scaling algorithm based on the space-time method has not yet been implemented within the NAO framework.

In previous works, we developed a low-scaling algorithm for RPA correlation-energy calculations for periodic systems with NAO basis sets \cite{ShiR24}. This algorithm was implemented in a standalone library LibRPA \cite{ShiR25,librpa-repo}, which has been interfaced with NAO-based DFT codes.
The algorithm was based on a real-space imaginary-time representation of the density response function $\chi^0$ together with the LRI approximation, and achieves a sub-quadratic scaling for the evaluation of $\chi^0$ with respect to the number of atoms in the unit cell.\cite{ShiR24}
In the present work, we extend this algorithm to $G^0W^0$ and present its implementation in LibRPA.
In particular, the low-scaling evaluation of the response function and self-energy is cast into a unified real-space tensor-contraction framework.
Benchmark calculations show that the low-scaling implementation reproduces the canonical $G^0W^0$ results in FHI-aims with high fidelity, while substantially reducing the scaling with respect to both $\bk$-point sampling and system size.
For the systems considered here, the overall scaling with system size is reduced to approximately $\mathcal{O}(N^{2.7})$, and good strong scaling is obtained up to the order of $10^4$ CPU cores.
Although the present implementation is formulated for NAOs, the underlying low-scaling algorithm is not restricted to NAOs and can in principle be applied in any localized atomic-orbital framework.

The remainder of this paper is organized as follows.
Sec.~\ref{sec:theory} recapitulates the fundamental formalism of $GW$ in a localized NAO basis and introduces the theoretical ingredients used in the present work.
Sec.~\ref{sec:implementation} describes the implementation details, including the optimized tensor-contraction algorithms, the treatment of the Coulomb-matrix singularity at the $\Gamma$ point, the overall workflow, and the computational setup used in this work.
Sec.~\ref{sec:results} presents benchmarks of the implementation in terms of accuracy and efficiency against the established canonical approach.
Finally, concluding remarks are given in Sec.~\ref{sec:conclude}.

\section{Theory and Methods}\label{sec:theory}

\subsection{The \textit{GW} method}\label{ssec:gw}
In this section, we recapitulate the essentials of the $GW$ method and its non-self-consistent scheme, i.e., the $G^0W^0$ approach.
The dynamical self-energy operator $\Sigma$ lies in the core of the \textit{GW} method.
In the imaginary time domain, it is formulated as
\begin{equation}\label{eq:gw-sigma}
  \Sigma(\bx,\bx',\ii\tau) = - G(\bx,\bx',\ii\tau) W(\bx,\bx', \ii\tau)
\end{equation}
where \(G\) is the interacting Green's function, $W$ the screened Coulomb interaction, and $\bx$ and $\bx'$ the electronic coordinates, including spatial and spin degrees of freedom.
The screened Coulomb interaction $W$ is usually calculated in the frequency domain by
\begin{equation}\label{eq:gw-w}
  W(\bx,\bx',\ii \omega) = v(\bx,\bx') +
  \int\dd{\bx_1}\dd{\bx_2} v(\bx,\bx_1) P^{\mathrm{RPA}}(\bx_1,\bx_2,\ii \omega) W(\bx_2,\bx',\ii \omega)
\end{equation}
where $v$ is the bare Coulomb interaction and $P$ is the polarization function or irreducible polarizability under the random-phase approximation (RPA), computed by Fourier transforming its time-domain counterpart
\begin{equation}\label{eq:gw-p}
  P^{\mathrm{RPA}}(\bx,\bx',\ii\tau) = G(\bx,\bx',\ii\tau) G(\bx',\bx,-\ii\tau)
\end{equation}
In principle, the quantities in Eqs.~\eqref{eq:gw-sigma}, \eqref{eq:gw-w}, and \eqref{eq:gw-p} should be expressed in terms of spin orbitals.
In this work, spin-orbit coupling (SOC) is neglected, so that the two spin channels are decoupled and both $\Sigma$ and $G$ are diagonal in spin space.
Furthermore, we limit the following discussion to the spin-restricted case, in which the two spin channels are equivalent, so that the explicit spin dependence can be omitted for notational simplicity.
Generalizations to spin-polarized and SOC cases are straightforward, and our code already supports both spin-polarized and SOC-included $GW$ calculations. Implementation and results for $GW$ calculations with SOC will be reported in a future publication.

In practice, $GW$ calculation begins with the Green's function of a reference system under a certain mean-field approximation and obtains the interacting Green's function by iteratively solving Eqs.~\eqref{eq:gw-sigma}, \eqref{eq:gw-w}, and \eqref{eq:gw-p} along with the Dyson equation until convergence is reached.
However, fully self-consistent $GW$ is not only computationally costly, but it has been shown to significantly overestimate quasi-particle band gaps of typical semiconductors.
Instead, the non-self-consistent $G^0W^0$ method provides a relatively cheap yet accurate approach for the quasi-particle band structure.
The initial Green's function at imaginary time is written as
\begin{equation}\label{eq:g0w0-g0}
G^0(\br,\br',\ii\tau) = \sum_{n\bk} \ee^{-\left(\epsilon_{n\bk} - \mu\right)\tau}  \left[ -\theta(\tau) (1-f_{n\bk}) + \theta(-\tau) f_{n\bk} \right] \psi_{n\bk}(\br) \psi^{*}_{n\bk}(\br')
\end{equation}
where $\epsilon_{n\bk}$, $\psi_{n\bk}$ and $f_{n\bk}$ are the orbital energy, wave function and occupation number of one-electron state $\ket{n\bk}$ of the reference system, respectively. $\theta(x)$ is the Heaviside step function.
Further taking into account that $P^{\mathrm{RPA}}$ is equal to the independent particle response function $\chi^0$, we rewrite Eqs.~\eqref{eq:gw-p} and \eqref{eq:gw-w} in $G^0W^0$ as
\begin{equation}\label{eq:g0w0-chi0}
 P^{\mathrm{RPA}}(\br,\br',\ii\tau) =  \chi^0(\br,\br',\ii\tau) = G^0(\br,\br',\ii\tau) G^0(\br',\br,-\ii\tau)
\end{equation}
and 
\begin{equation}\label{eq:g0w0-w0-dm-realspace}
  W^0(\br,\br',\ii \omega) = 
  \int\dd{\br''} \varepsilon^{-1}(\br,\br'', \ii \omega) v(\br'',\br'),
\end{equation}
where $\varepsilon$ is the RPA microscopic dielectric function
\begin{equation}\label{eq:g0w0-dm-realspace}
\varepsilon(\br,\br', \ii \omega) = \delta(\br,\br') - \int\dd{\br''} v(\br,\br'')\chi^0(\br'',\br', \ii \omega)\, .
\end{equation}
Then, the screened Coulomb interaction $W^0$ calculated in \eqref{eq:g0w0-w0-dm-realspace} is Fourier-transformed from the imaginary frequency to the imaginary time domain, $W(\br,\br', \ii \omega) \rightarrow W(\br,\br', \ii \tau)$.  Further using the expression for non-interacting Green function $G^0$ in Eq.\eqref{eq:g0w0-g0}, the $G^0W^0$ self-energy in the imaginary time domain can be obtained as a simple product,
\begin{equation}\label{eq:g0w0-sigma}
\Sigma(\br,\br',\ii\tau) = - G^0(\br,\br',\ii\tau) W^0(\br,\br', \ii\tau)\, .
\end{equation}
Afterwards, the $G^0W^0$ self-energy is Fourier-transformed back from the (imaginary) time to the frequency domain,
and then analytically continued from the imaginary frequency axis to the real axis.

The $G^0W^0$ quasi-particle energy of state $\ket{n\bk}$, $\epsilon_{n\bk}^{\QP}$, is obtained by solving the non-linear quasi-particle equation of $\omega$
\begin{equation}\label{eq:quasi-particle-equation}
\omega = \epsilon_{n\bk} - v^{\xc}_{n\bk} + \mel{n\bk}{\hat{\Sigma}(\omega)}{n\bk}
\end{equation}
where $v^{\xc}_{n\bk}$ is the exchange-correlation potential obtained from the mean-field reference calculation, and $\hat{\Sigma}(\omega)$ the self-energy operator in real frequency. The formalism is generally known as the space-time approach \cite{RojasH95}.
We note that different papers adopt different phase conventions in the imaginary-time formalism;
see Eqs.~\eqref{eq:g0w0-g0}, \eqref{eq:g0w0-chi0}, and \eqref{eq:g0w0-sigma}.
These differences stem from the chosen Fourier-transform convention connecting the imaginary-time and imaginary-frequency representations.

The original space-time formalism relies on fast Fourier transforms (FFTs) and typically requires rather dense real-space and imaginary-time grids to converge the results.\cite{RiegerM99,SteinbeckL00,LiuP16,FoersterD21}
As a consequence, substantial memory is needed to store the real-space Green’s function and the resulting response functions and self-energies.
To mitigate these bottlenecks, several improvements have been proposed. On the time-frequency side, Kaltak and co-workers introduced optimized non-uniform grids and efficient transforms, which have since been adopted in both RPA and \textit{GW} calculations.\cite{KaltakM14,KaltakM14Si,LiuP16}
On the real-space representation side, one may avoid an explicit grid by working in a localized basis.\cite{BlumV09}
While this sacrifices the pointwise multiplicative structure of $\chi$ and $\Sigma$ in real space and turns the corresponding convolutions into matrix multiplications,
the sparsity in matrices can be exploited to accelerate the calculation (see Sec.~\ref{ssec:gw-with-nao} for details).

\subsection{Numerical atomic orbital formalism}\label{ssec:nao}
To make the equations formulated in Sec. \ref{ssec:gw} computable,
the real-space quantities have to be projected onto a linear space spanned by a finite set of basis functions.
Popular choices for basis functions include plane waves (PW), linearized augmented plane waves (LAPW) and Gaussian-type orbitals (GTOs).
In this work, we adopt the NAOs, a type of localized basis functions with values of their radial part 
tabulated on real-space grids.
An NAO basis function centered on atom $I$ in the unit cell marked by a real space vector $\bR$ is written as
\begin{equation}\label{eq:NAO_basis}
\varphi_{i\bR}(\br) \equiv \varphi_{i}(\br-\mathbf{t}_{I}-\bR)
\end{equation}
where $\mathbf{t}_{I}$ denotes the position vector of atom $I$ within the unit cell. The collective index $i$ uniquely identifies the basis function, i.e., the atom, the associated radial function, and the angular channel.
The wave function with band index $n$ and momentum vector $\bk$ is then expanded in terms of the corresponding Bloch sums of NAOs, $\varphi_{i\bk}$, as
\begin{equation}\label{eq:wfc}
\begin{aligned}
\psi_{n\bk}(\br) = \sum_i c^i_{n\bk} \varphi_{i\bk}(\br)
= \invsqNk \sum_i c^i_{n\bk} \sum_{\bR} \ee^{\ii\bk\cdot\bR} \varphi_{i\bR}(\br)
\end{aligned}
\end{equation}
where $c^i_{n\bk}$ are the expansion coefficients and $N_{\bk}$ the number of $\bk$ points sampled in the first Brillouin zone (BZ), equivalent in our implementation to the number of unit cells in the Born-von K\'{a}rm\'{a}n (BvK) supercell.

\subsection{Localized resolution of identity technique for the NAOs in periodic systems}
The resolution of identity (RI) approach refers to expanding the products of wave functions in terms of
a set of auxiliary basis functions (ABFs).
This is equivalent to expanding the products of NAO basis functions by the same set of ABFs.
In periodic systems, such an expansion is formally given by \cite{RenX21}
\begin{equation}\label{eq:ri}
\varphi_{i\bR}(\br)\varphi_{j\bR'}(\br) \approx
\sum_{\mu\bR_1} C^{\mu(\bR_1)}_{i(\bR), j(\bR')} P_{\mu\bR_1}(\br)
\end{equation}
where $P_{\mu\bR_1}$ is the ABF centered on atom $\cMu$ in a unit cell specified by lattice vector $\bR_1$, defined similarly to the orbital basis in Eq. \eqref{eq:NAO_basis}, i.e.,
\begin{equation}\label{eq:NAO_abf}
P_{\mu\bR}(\br) \equiv P_{\mu}(\br - \mathbf{t}_{\cMu} - \bR).
\end{equation}
$C^{\mu(\bR_1)}_{i(\bR), j(\bR')}$ are usually called the RI expansion coefficients or triple coefficients%
.
$C^{\mu(\bR_1)}_{i(\bR), j(\bR')}$ satisfies the translational symmetry and the commutative rule
\begin{equation}\label{eq:C-period-commutative}
C^{\mu(\bR_1)}_{i(\bR), j(\bR')} = C^{\mu(\mathbf{0})}_{i(\bR-\bR_1), j(\bR'-\bR_1)}
= C^{\mu(\mathbf{0})}_{j(\bR'-\bR_1),i(\bR-\bR_1)}
\end{equation}
The ABFs share the same function form as the single-particle NAO basis, but use different radial functions.
These radial functions can be provided \textit{a priori} or generated on the fly from the single-particle NAO basis set. A robust prescription for on-the-fly generation of ABFs has been provided in Refs.~\citenum{RenX12-RI-NAO,IhrigA15}.

In principle, the index $\mu$ in Eq.\eqref{eq:NAO_abf} runs over ABFs on all atoms $\cMu$ for each one-electron basis pair $ij$. The resulting expression for the response function in localized basis framework therefore scales as $\mathcal{O}(N_a^6)$ with $N_a$ the number of atoms (see discussions in Sec. \ref{ssec:gw-with-nao}). This corresponds to the global RI scheme.
In practice, only atom triplets $\langle IJ\cMu\rangle$ with \textit{non-zero} RI coefficients need to be retained, which reduces the formal scaling to $\mathcal{O}(N_a^4)$. The reduction arises because, for a given atom-pair block $\langle IJ\rangle$, the auxiliary-center index $\cMu$ is effectively constrained by the sparsity pattern of the RI coefficients, rather than running independently over all atoms in the system.
The actual sparsity of RI coefficients tensor depends on the metric with which they are computed.\cite{RenX12-RI-NAO} The Coulomb metric, suitable for correlation-type calculations, results in non-zero coefficients even with $\cMu$ far away from the $ij$ pair due to the long-range nature of the electrostatic interaction. Consequently, although the formal scaling is quartic, the prefactor remains very large, and this asymptotic behavior can only be reached for very large systems.\cite{RenX12-RI-NAO,ForsterA20,WilhelmJ21}

Improvements from the standard global RI scheme have been proposed.
These include RI schemes based on attenuated or truncated Coulomb metrics,\cite{LuenserA17,WilhelmJ21} as well as separable RI schemes.\cite{DucheminI21,DelesmaFA24}
Among them, the localized variant of RI assumes that the ABFs must be located on the same atom
as either of the two basis functions, i.e.,\cite{IhrigA15}
\begin{equation}\label{eq:lri}
\begin{aligned}
\varphi_{i\bR}(\br)\varphi_{j\bR'}(\br) &\approx \sum_{\mu\in I} C^{\mu(\bR)}_{i(\bR), j(\bR')} P_{\mu\bR}(\br)
+\sum_{\mu\in J} C^{\mu(\bR')}_{i(\bR), j(\bR')} P_{\mu\bR'}(\br) \\
&= \sum_{\mu\in I} C^{\mu(\mathbf{0})}_{i(\mathbf{0}), j(\bR'-\bR)} P_{\mu\bR}(\br)
+\sum_{\mu\in J} C^{\mu(\mathbf{0})}_{j(\mathbf{0}), i(\bR-\bR')} P_{\mu\bR'}(\br). 
\end{aligned}
\end{equation}
The second equality is derived from Eq. \eqref{eq:C-period-commutative}. The notation $\mu\in I$ ($\mu\in J$) indicates that the ABF index $\mu$ only runs over those residing on the atom $I$ ($J$).
This local approximation significantly reduces the number of triple coefficients to compute and store,
which facilitates the design of low-scaling algorithms for advanced \textit{ab initio} methods.
This approach is referred to as local RI (LRI), also known as pair-atom density fitting (PADF)
\cite{Merlot/etal:2013,Wirz/etal:2017} or concentric atomic density fitting (CADF) \cite{Hollman/Schaefer/Valeev:2014,Wang/Lewis/Valeev:2020} in the literature.
The LRI approach has been successfully applied in hybrid functional \cite{Levchenko/etal:2015,Lin/Ren/He:2020,LinP21JCTC,Kokott/etal:2024,Lin/etal:2025}, RPA \cite{Spadetto/etal:2023,ShiR24,ShiR25}, and $G^0W^0$ 
calculations \cite{ForsterA20,RenX21}, as well as RPA force calculations for molecules \cite{Tahir/etal:2022,Tahir/etal:2025}.
The LRI scheme lays the foundation of our low-scaling implementation, which we discuss extensively in the following section.

\subsection{\textit{GW} formulation in NAOs using LRI}\label{ssec:gw-with-nao}

\subsubsection{Response function in ABF representation}
Here we present the working $GW$ equations under the representation of NAO basis sets and
the LRI approximation. To begin with, the non-interacting Green's function in Eq. \eqref{eq:g0w0-g0} is first expanded in terms of
NAOs. Using Eqs.~\eqref{eq:g0w0-g0} and \eqref{eq:wfc}, we obtain
\begin{equation}\label{eq:g0-nao}
G^0(\br,\br',\ii\tau) =
\sum_{ij} \sum_{\bR,\bR'} \varphi_{i\bR}(\br) G^0_{ij}(\bR'-\bR,\ii\tau) \varphi_{j\bR'}(\br')
\end{equation}
with
\begin{align}
G^0_{ij}(\bR,\ii\tau) =& \invNk \sum_{n\bk} \ee^{-\ii\bk\cdot\bR} c_{n\bk}^i c_{n\bk}^{j*} \Xi_{n\bk}(\tau) \\
\Xi_{n\bk}(\tau) =& \ee^{-(\epsilon_{n\bk}-\mu)\tau}
\left[\theta(\tau)(1-f_{n\bk}) - \theta(-\tau)f_{n\bk} \right]\,.
\end{align}
By inserting Eq. \eqref{eq:g0-nao} into Eq. \eqref{eq:g0w0-chi0}, the density response function $\chi^0$ can be written as
\begin{equation}
\begin{aligned}
\chi^0(\br,\br',\ii\tau) =& \sum_{ij} \sum_{\bR,\bR'} \varphi_{i\bR}(\br) \varphi_{j\bR'}(\br') G^0_{ij}(\bR'-\bR,\ii\tau) \times \\
&\qquad \sum_{kl} \sum_{\bR_1,\bR_2} \varphi_{l\bR_2}(\br') \varphi_{k\bR_1}(\br) G^0_{lk}(\bR_1-\bR_2,-\ii\tau)
\end{aligned}
\end{equation}
Since we consider only the one-shot $G^0W^0$ approach in this work, we suppress the superscript on $G^0$ for brevity. 
The products of NAO basis functions are then expanded in terms of ABFs under the LRI approximation [Eq. \eqref{eq:lri}], giving rise to
\begin{equation}\label{eq:chi0_lri}
\begin{aligned}
\chi^0(\br,\br',\ii\tau)
=& \sum_{ijkl} \sum_{\bR,\bR',\bR_1,\bR_2} \left[ \sum_{\mu\in I} C^{\mu(\mathbf{0})}_{i(\mathbf{0}), k(\bR_1-\bR)} P_{\mu\bR}(\br)
+\sum_{\mu\in K} C^{\mu(\mathbf{0})}_{k(\mathbf{0}), i(\bR-\bR_1)} P_{\mu\bR_1}(\br) \right] \\
&\qquad\qquad\left[ \sum_{\nu\in J} C^{\nu(\mathbf{0})}_{j(\mathbf{0}), l(\bR_2-\bR')} P_{\nu\bR'}(\br')
+\sum_{\nu\in L} C^{\nu(\mathbf{0})}_{l(\mathbf{0}), j(\bR'-\bR_2)} P_{\nu\bR_2}(\br') \right] \\
&\qquad\qquad\qquad G_{ij}(\bR'-\bR,\ii\tau) G_{lk}(\bR_1-\bR_2,-\ii\tau) \, .
\end{aligned}
\end{equation}
We note that the constraint in the $\mu$- and $\nu$-summation is absent in global RI, resulting in more sophisticated scaling or enhanced prefactor.
By changing the indices and using the translational symmetry of RI coefficients and Green's function, Eq.~\eqref{eq:chi0_lri} can be rewritten into a compact form as
\begin{equation}
\chi^0(\br,\br',\ii\tau) = \sum_{\mu\nu}\sum_{\bR\bR'} P_{\mu\bR}(\br) \chi^0_{\mu\nu}(\bR'-\bR,\ii\tau) P_{\nu\bR'}(\br')
\end{equation}
where
\begin{equation}\label{eq:chi0-mn}
\begin{aligned}
\chi^0_{\mu\nu}(\bR,\ii\tau) = \sum_{i \in \cMu, j\in\cNu} & \sum_{k\bR_1, l\bR_2} C^{\mu(\mathbf{0})}_{i(\mathbf{0}),k(\bR_1)} C^{\nu(\bR)}_{j(\bR),l(\bR_2)} \\
&\left[ G_{ij}(\bR,\ii\tau) G_{lk}(\bR_1-\bR_2,-\ii\tau) +
G_{il}(\bR_2,\ii\tau) G_{jk}(\bR_1-\bR,-\ii\tau) + \right. \\
&\left.
G_{ji}(-\bR,-\ii\tau) G_{kl}(\bR_2-\bR_1,\ii\tau) +
G_{li}(-\bR_2,-\ii\tau) G_{kj}(\bR-\bR_1,\ii\tau) \right].
\end{aligned}
\end{equation}
An efficient algorithm to compute Eq.~\eqref{eq:chi0-mn} has been discussed in our previous work for RPA correlation energy calculations.\cite{ShiR24}
In Sec.~\ref{ssec:contraction}, we recapitulate the essentials of the algorithm and further demonstrate that it shares a similar algebraic structure with the self-energy operator; hence, both can be computed in a unified tensor contraction framework.

\subsubsection{Screened Coulomb interaction}
To proceed to the construction of screened Coulomb interaction,
the real-space imaginary-time response function matrix $\chi^0_{\mu\nu}(\bR,\ii\tau)$ needs to be transformed to $\bk$ space and imaginary-frequency domain.
In practice, the transformation is done on a discretized time and frequency grids
via the cosine transform \cite{KaltakM14,LiuP16} (see Appendix \ref{app:ft}),
\begin{equation}
\chi^0_{\mu\nu}(\bR,\ii\omega_k) = \sum_{j=1}^{N_\tau} \gamma_{kj} \cos\left(\omega_k \tau_j\right) \chi^0_{\mu\nu}(\bR,\ii\tau_j)
\end{equation}
where $N_\tau$ is the number of imaginary time points, and then Fourier transformed to reciprocal space
\begin{equation}\label{eq:chi0-mn-qw}
\chi^0_{\mu\nu}(\bq,\ii\omega_k) = \sum_{\bR} \ee^{\ii\bq\cdot\bR} \chi^0_{\mu\nu}(\bR,\ii\omega_k).
\end{equation}
In this work, we employ the optimized minimax time/frequency grids implemented in the GreenX library.\cite{AziziM23} The transformation coefficients $\{\gamma_{kj}\}$ are generated on the fly with the tabulated grid points $\{\omega_k\}$ and $\{\tau_j\}$.

The screened Coulomb interaction in Eq.~\eqref{eq:g0w0-w0-dm-realspace} is represented by the ABFs as
\begin{equation}\label{eq:w-mn-qw}
W_{\mu\nu}(\bq,\ii\omega) = \int\dd{\br}\dd\br' P^{*}_{\mu\bq}(\br) W(\br,\br',\ii\omega) P_{\nu\bq}(\br')
\end{equation}
where $P_{\mu\bq}(\br)=\sum_{\bR} e^{i\bq \cdot \bR} P_{\mu \bR}(\br)$ is the Bloch sum of the ABFs.
Combined with the Coulomb matrix
\begin{equation}\label{eq:v-mn-q}
V_{\mu\nu}(\bq) = \sum_{\bR} \ee^{\ii\bq\cdot\bR} V_{\mu\nu}(\bR) = \sum_{\bR} \ee^{\ii\bq\cdot\bR} \iint\dd{\br}\dd{\br'} P_{\mu\mathbf{0}}(\br) v(\abs{\br-\br'}) P_{\nu\bR}(\br') \,,
\end{equation}
where $v(r) = 1/r$ is the bare Coulomb kernel,
Eq.~\eqref{eq:gw-w} can be reformulated in matrix form as
\begin{equation}\label{eq:w-dyson-matrix}
W_{\mu\nu}(\bq,\ii\omega) = V_{\mu\nu}(\bq) + \sum_{\mu'\nu'} V_{\mu\mu'}(\bq) \chi^0_{\mu'\nu'}(\bq,\ii\omega) W_{\nu'\nu}(\bq,\ii\omega)\, .
\end{equation}
\JCTC{%
We note that, for three-dimensional systems, the lattice sum in Eq.~\eqref{eq:v-mn-q} is not absolutely convergent for matrix elements between ABFs with nonvanishing multipole moments when \(l(\mu)+l(\nu)\le 2\), where \(l(\mu)\) is the angular momentum quantum number associated with \(P_{\mu}\).\cite{StolarczykLZ82}
This behavior is directly related to the singular nature of the Coulomb matrix at \(\bq=0\). 
The technique used in this work to handle the Coulomb-matrix singularity is discussed in Sec.~\ref{ssec:finite-size}.
}%

It is convenient to introduce the symmetrized dielectric matrix
\begin{equation}\label{eq:symeps}
\tilde{\varepsilon}_{\mu\nu}(\bq,\ii\omega) = \delta_{\mu\nu} - \sum_{\mu'\nu'} V^{1/2}_{\mu\mu'}(\bq) \chi^0_{\mu'\nu'}(\bq,\ii\omega) V^{1/2}_{\nu'\nu}(\bq)
\end{equation}
and then
\begin{equation}\label{eq:w0-mu-qw-symeps}
\begin{aligned}
W_{\mu\nu}(\bq,\ii\omega) =& \sum_{\mu'\nu'} V^{1/2}_{\mu\mu'} \left[\tilde{\varepsilon}^{-1}(\bq,\ii\omega)\right]_{\mu'\nu'} V^{1/2}_{\nu'\nu}\, .
\end{aligned}
\end{equation}
In practice, the correlation part of screened Coulomb is obtained by subtracting the bare Coulomb matrix from Eq. \eqref{eq:w0-mu-qw-symeps}
\begin{equation}
\Wc_{\mu\nu}(\bq,\ii\omega) = W_{\mu\nu}(\bq,\ii\omega) - V_{\mu\nu}(\bq)\,,
\end{equation}
and its counterpart in imaginary-time domain is obtained by performing an inverse cosine transform
\begin{equation}\label{eq:w0-mu-qt}
\Wc_{\mu\nu}(\bq,\ii\tau_j) = \sum_{k=1}^{N_{\omega}} \xi_{jk} \cos\left(\tau_j \omega_k\right) \Wc_{\mu\nu}\left(\bq, \ii\omega_k\right)\,
\end{equation}
where $N_{\omega}$ is the number of imaginary frequency points ($N_{\omega}=N_{\tau}$ in the present work) and $\{\xi_{jk}\}$ the transformation coefficients.

\subsubsection{Self-energy matrix in NAO basis on imaginary time/frequency axis}\label{sssec:sigma-nao}
To solve the quasi-particle equation [Eq.~\eqref{eq:quasi-particle-equation}], we compute the self-energy on imaginary frequency axis and then analytically continue to the real axis. The diagonal self-energy matrix elements for KS states are obtained by
\begin{equation}\label{eq:Sigma_XC_KS}
\Sigma_{nn}(\bk,\ii\omega) = \mel{n\bk}{\hat{\Sigma}(\ii\omega)}{n\bk} = \sum_{ij} c^{i*}_{n\bk} c^{j}_{n\bk} \Sigma_{ij}(\bk,\ii\omega) 
\end{equation}
with
\begin{equation}
\Sigma_{ij}(\bk,\ii\omega) = \sum_{\bR} \ee^{\ii\bk\cdot\bR} \Sigma_{ij}(\bR,\ii\omega)
\end{equation}
where $\Sigma_{ij}(\bR,\ii\omega)$ is the real-space self-energy matrix in NAO basis.
In practice, $\Sigma_{ij}(\bR,\ii\omega)$ is separated into the exchange and correlation contributions
\begin{equation}\label{eq:Sigma_XC_ij}
\Sigma_{ij}(\bR,\ii\omega) = \Sigma^\mathrm{x}_{ij}(\bR) + \Sigma^{\mathrm{c}}_{ij}(\bR,\ii\omega)
\end{equation}
The time-independent term $\Sigma^{\rm x}_{ij}(\bR)$ is the non-local exact-exchange potential matrix. Under
the LRI approximation \cite{IhrigA15,LinP21JCTC}, it is given by
\begin{equation}\label{eq:vexx-R}
\begin{aligned}
\Sigma^\mathrm{x}_{ij}(\bR) =& - \sum_{k\bR_1}\sum_{l\bR_2} D_{kl}(\bR_2-\bR_1) \times \left[\sum_{\mu\in I}\sum_{\nu\in J} C^{\mu(\mathbf{0})}_{i(\mathbf{0}),k(\bR_1)} V_{\mu\nu}(\bR) C^{\nu(\bR)}_{j(\bR),l(\bR_2)} \right. \\
& + \sum_{\mu\in I}\sum_{\nu\in L} C^{\mu(\mathbf{0})}_{i(\mathbf{0}),k(\bR_1)} V_{\mu\nu}(\bR_2) C^{\nu(\bR_2)}_{j(\bR),l(\bR_2)}
+ \sum_{\mu\in K}\sum_{\nu\in J} C^{\mu(\bR_1)}_{i(\mathbf{0}),k(\bR_1)} V_{\mu\nu}(\bR-\bR_1) C^{\nu(\bR)}_{j(\bR),l(\bR_2)}\\
& + \left.
\sum_{\mu\in K}\sum_{\nu\in L} C^{\mu(\bR_1)}_{i(\mathbf{0}),k(\bR_1)} V_{\mu\nu}(\bR_2-\bR_1) C^{\nu(\bR_2)}_{j(\bR),l(\bR_2)} \right] \\
\end{aligned}
\end{equation}
with the real-space density matrix
\begin{equation}
D_{ij}(\bR) = \frac{1}{N_\bk} \sum_{n\bk} \ee^{-\ii\bk\cdot\bR} D_{ij}(\bk) = \frac{1}{N_\bk} \sum_{n\bk} \ee^{-\ii\bk\cdot\bR} f_{n\bk} c_{n\bk}^{i} c_{n\bk}^{j*}.
\end{equation}
Efficient algorithms to compute Eq.~\eqref{eq:vexx-R} have been extensively investigated in previous works.\cite{IhrigA15,LinP21JCTC,Kokott/etal:2024,Lin/etal:2025,CaoY2025}

The evaluation of correlation self-energy $\Sigma^{\mathrm{c}}_{ij}(\bR,\ii\omega)$ is the central and the most time-consuming component of the \textit{GW} method.
In our low-scaling algorithm, we compute the imaginary-time self-energy $\Sigma^{\mathrm{c}}_{ij}(\bR,\ii\tau)$. With the LRI approximation, it can be derived analogously to the exchange counterpart [Eq.~\eqref{eq:vexx-R}] as
\begin{equation}\label{eq:sigmac-ij-Rt}
\begin{aligned}
\Sigma^{\mathrm{c}}_{ij}(\bR,\ii\tau) =& -\sum_{k\bR_1}\sum_{l\bR_2} G_{kl} (\bR_2 - \bR_1,\ii\tau) \times \\
& \left[\sum_{\mu\in I}\sum_{\nu\in J} C^{\mu(\mathbf{0})}_{i(\mathbf{0}),k(\bR_1)} \Wc_{\mu\nu}(\bR,\ii\tau) C^{\nu(\bR)}_{j(\bR),l(\bR_2)}\right. + \\
& \,\sum_{\mu\in K} \sum_{\nu\in J} C^{\mu(\bR_1)}_{i(\mathbf{0}),k(\bR_1)} \Wc_{\mu\nu}(\bR-\bR_1,\ii\tau) C^{\nu(\bR)}_{j(\bR),l(\bR_2)} + \\
& \,\sum_{\mu\in I}\sum_{\nu\in L} C^{\mu(\mathbf{0})}_{i(\mathbf{0}),k(\bR_1)} \Wc_{\mu\nu}(\bR_2,\ii\tau) C^{\nu(\bR_2)}_{j(\bR),l(\bR_2)} + \\
& \,\left.\sum_{\mu\in K} \sum_{\nu\in L} C^{\mu(\bR_1)}_{i(\mathbf{0}),k(\bR_1)} \Wc_{\mu\nu}(\bR_2-\bR_1,\ii\tau) C^{\nu(\bR_2)}_{j(\bR),l(\bR_2)}\right]
\end{aligned}
\end{equation}
where the real-space 2-center screened Coulomb matrix is obtained via Fourier transforming  [Eq.\eqref{eq:w0-mu-qt}]
\begin{equation}
W_{\mu\nu}(\bR,\ii\tau) = \frac{1}{N_\bk} \sum_{\bq} \ee^{-\ii\bq\cdot\bR} W_{\mu\nu}(\bq,\ii\tau)\, .
\end{equation}
The correlation self-energy in Eq.~\eqref{eq:sigmac-ij-Rt} shares the algebraic structure with the exchange counterpart in Eq.~\eqref{eq:vexx-R}, except for using Green's function instead of the density matrix, and screened Coulomb interaction instead of the bare one.
Thus, essentially, the numerical technique applied to Eq.~\eqref{eq:vexx-R} also works for Eq.~\eqref{eq:sigmac-ij-Rt}, which will be discussed in detail in Sec.~\ref{ssec:contraction}.
The self-energy in frequency domain is constructed from both cosine and sine transforms\cite{LiuP16}
\begin{equation}
\Sigma^{\mathrm{c}}_{ij}(\bR,\ii\omega_k) = \sum_{j=1}^{N_\tau} \gamma_{kj} \cos\left(\omega_k \tau_j\right) \Sigma^{\mathrm{c,even}}_{ij}(\bR,\ii\tau_j) + 
\ii \sum_{j=1}^{N_\tau} \lambda_{kj} \sin\left(\omega_k \tau_j\right) \Sigma^{\mathrm{c,odd}}_{ij}(\bR,\ii\tau_j)
\end{equation}
where
\begin{subequations}
\begin{equation}
\Sigma^{\mathrm{c,even}}_{ij}(\bR,\ii\tau) = \frac{1}{2}\left[\Sigma^{\mathrm{c}}_{ij}(\bR,\ii\tau) + \Sigma^{\mathrm{c}}_{ij}(\bR,-\ii\tau)\right]
\end{equation}
\begin{equation}
\Sigma^{\mathrm{c,odd}}_{ij}(\bR,\ii\tau) = \frac{1}{2}\left[\Sigma^{\mathrm{c}}_{ij}(\bR,\ii\tau) - \Sigma^{\mathrm{c}}_{ij}(\bR,-\ii\tau)\right]
\end{equation}
\end{subequations}
and $\{\lambda_{kj}\}$ are the sine transformation coefficients.
The self-energy at negative imaginary time $\Sigma^{\mathrm{c}}(-\ii\tau)$ can be also computed with Eq.~\eqref{eq:sigmac-ij-Rt}
and $\Wc(\ii\tau)$ at positive time, exploiting the even parity of $\Wc$ in time domain.

\subsubsection{Analytic continuation with Pad\'e approximant}\label{sssec:ac}
To obtain the self-energy along real frequency axis for the quasi-particle equation,
we utilize the Pad\'e approximant to analytically continue the self-energy at imaginary frequencies [Eq.~\eqref{eq:Sigma_XC_KS}].
The procedure of analytical continuation is essentially to approximate the unknown function $f(z)$ with a chosen analytic form and uniquely define the function by establishing the necessary coefficients from a set of known points $\{z_n, f_n\}$, such that $f_n = f(z_n)$. In our case, the known data points are $\{\ii\omega_n, \Sigma^{\mathrm{c}}(\ii\omega_n)\}$ for each Kohn-Sham state.
For the $N$-point Pad\'e approximant of complex variable $z$,
\begin{equation}\label{eq:nth-pade-approximant}
\begin{aligned}
f(z) \approx P_{N}(z) = \cfrac{a_1}{1 +
\cfrac{a_2 \left(z-z_1\right)}{1+ \ldots \frac{a_N\left(z-z_{N-1}\right)}{1 + \left(z-z_{N}\right) g_{N+1}(z) }}}
\end{aligned}
\end{equation}
where $\{a_1, a_2, \ldots, a_N\}$ are the coefficients to be determined and
\begin{equation}
\begin{aligned}
g_{n}(z) ~=&~ \frac{a_n}{1 + \left(z-z_{n}\right) g_{n+1}(z)}, \forall n = 1, 2, \ldots, N \\
g_{N+1}(z) ~=&~ 0.
\end{aligned}
\end{equation}
Starting from $g_{1}(z) = P_N(z)$,
the Thiele's reciprocal difference method\cite{BakerG75,VidbergH77} provides an efficient way to determine the coefficients using $P_N(z_n) = f_n$ and
\begin{equation}
\begin{aligned}
g_n(z) =& \frac{g_{n-1}(z_{n-1}) - g_{n-1}(z)}{(z-z_{n-1})g_{n-1}(z)}, \quad \forall n\ge 2.
\end{aligned}
\end{equation}
The coefficients are simply $a_n = g_n(z_n)$.%

\section{Implementation}\label{sec:implementation}

In this section, we describe the main implementation aspects of the present low-scaling $G^0W^0$ approach.
We begin with the efficient tensor-contraction algorithms for the response function and self-energy, which constitute the computational core of the method.
Next, we describe the treatment of the Coulomb-matrix singularity at the $\Gamma$ point, and then summarize the overall computational workflow and the technical details adopted in this work.

\subsection{Efficient tensor contraction for response function and self-energy}\label{ssec:contraction}
We first discuss in detail our algorithm used to compute the real-space response function $\chi^0_{\mu\nu}(\bR,\ii\tau)$ and correlation self-energy $\Sigma^{\mathrm{c}}_{ij}(\bR,\ii\tau)$,
namely Eq.~\eqref{eq:chi0-mn} and Eq.~\eqref{eq:sigmac-ij-Rt}.
To facilitate the discussion, the matrix/tensor blocks in NAO basis are labeled by
the atoms on which the basis functions associated with the row and column indices reside.
Accordingly, we denote the atom-pair block of matrix $\mathbf{A}$ as $\mathbf{A}_{IJ} = \{A_{ij}|i\in I, j\in J\}$. In the real-space representation, we further combine atom indices with unit cell vectors to simplify the notation, {e.g.}, $(I,\mathbf{0}) \to I$, $(J,\bR) \to J$, $(K,\bR_1) \to K$, $(L,\bR_2) \to L$, when no ambiguity arises. Under this convention, the indices should be viewed as labels of atoms in the BvK cell rather than in the unit cell.
\JCTC{%
We note that the algorithms presented below can be viewed as periodic counterparts of the molecular formulation introduced in Ref.~\citenum{ForsterA20}, with the additional treatment of Bloch wave vectors and lattice periodicity.
}%

\subsubsection{Response function}\label{sssec:contraction-chi}
We first investigate the computation of blocks of real-space response function matrix $\boldsymbol{\chi}^0_{\cMu\cNu} = \{\chi^0_{\mu\nu}|\mu\in\cMu, \nu\in\cNu\}$ in Eq.~\eqref{eq:chi0-mn}. Using the atom-pair block notation and denoting the positive-(negative-)time Green's function as $G^{+}$ ($G^{-}$), we can rewrite Eq.~\eqref{eq:chi0-mn} as
\begin{equation}\label{eq:chi0-mn-ap-block}
\begin{aligned}
\boldsymbol{\chi}^0_{\cMu\cNu}(\ii\tau) = \sum_{I=\cMu, J=\cNu} \sum_{KL} \mathbf{C}^{\cMu}_{IK} & \left[ \mathbf{G}^{+}_{IJ} \mathbf{G}^{-}_{LK} +
\mathbf{G}^{+}_{IL} \mathbf{G}^{-}_{JK} +
\mathbf{G}^{-}_{JI} \mathbf{G}^{+}_{KL} +
\mathbf{G}^{-}_{LI} \mathbf{G}^{+}_{KJ} \right] \mathbf{C}^{\cNu}_{JL}
\end{aligned}
\end{equation}
where the matrix multiplication and tensor contraction are carried out along basis indices on recurring atoms. Apparently, the computation scales as $\mathcal{O}(N^4_a N^3_k)$ considering the loops over blocks $\expval{\cMu(\mathbf{0})\cNu(\bR)}$ and atom pairs $\expval{K(\bR_1)L(\bR_2)}$ in the summation, where $N_a$ is the number of atoms in the unit cell and $N_k$ is the number of k-points in the BZ sampling.
However, the sparsity of local RI coefficients tensor ($\mathbf{C}^{\cMu=I}_{IK}$, $\mathbf{C}^{\cNu=J}_{JL}$) ensures that atom $K(\bR_1)$ and $L(\bR_2)$ can only be taken from the neighborhoods of atoms $\cMu(\mathbf{0})$ and $\cNu(\bR)$, respectively. As a result, the summations over $K(\bR_1)$ and $L(\bR_2)$ saturate and only contribute to a prefactor as the system size increases, and the scaling is reduced to $\mathcal{O}(N^2_a N_k)$. The sparsity in RI coefficient tensors is evaluated by a filtering threshold $\eta_C$. Moreover, by filtering the Green's function matrix with a controllable threshold $\eta_G$, one can further reduce to sub-quadratic scaling with $N_a$ as reported in Ref. \citenum{ShiR24}.

\begin{figure}
\includegraphics[width=0.96\linewidth]{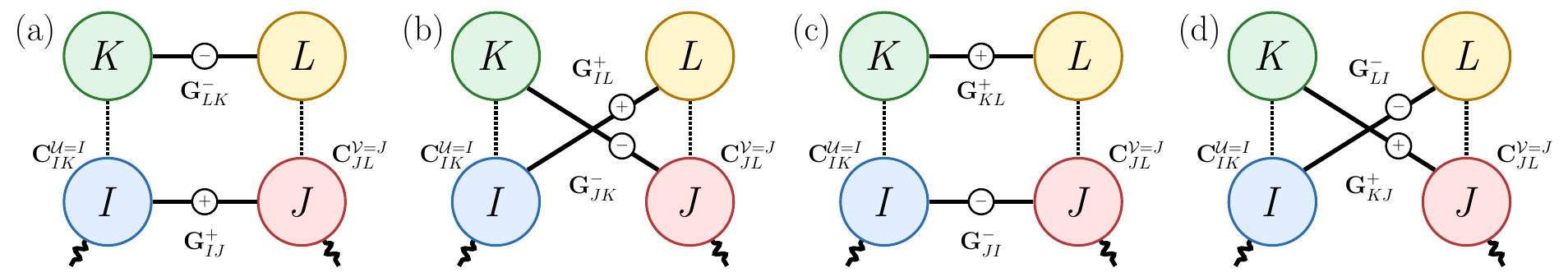}
\caption{\label{fig:chi0-tensor-contract-pattern}
  Schematic diagrams of the contributions from an atom quartet $\expval{IJKL}$ to the atom-pair block of the real-space imaginary-time response function $\chi^0_{\cMu=I,\cNu=J}(\ii\tau)$ in Eq.~\eqref{eq:chi0-mn-ap-block}.
  Here, $\mathbf{G}^{\pm} = \mathbf{G}(\pm\ii\tau)$.
  From left to right, the diagrams correspond to the four terms in Eq.~\eqref{eq:chi0-mn-ap-block}, respectively.
  The atom indices denote atoms in the BvK cell.
  The dashed lines indicate the intrinsic binding of atom pairs in the RI-coefficient tensor $\mathbf{C}^{\cMu}_{IK}$ under the LRI approximation.
  The solid straight lines represent one-electron NAO basis indices that are contracted during the computation.
  The auxiliary-basis indices are denoted by wavy lines and remain uncontracted for the response function.
}
\end{figure}

To implement Eq.~\eqref{eq:chi0-mn-ap-block}, a na\"ive loop over all atom quartets is computationally inefficient. In Ref.~\citenum{ShiR24}, we introduced an efficient algorithm for this purpose. Here we reformulate it in terms of tensor-contraction patterns.
The four contributions from each atom quartet $\expval{IJKL}$ to $\boldsymbol{\chi}^0_{\cMu=I,\cNu=J}$ are illustrated in Fig.~\ref{fig:chi0-tensor-contract-pattern}.
As can be seen, the first and third terms in Eq.~\eqref{eq:chi0-mn-ap-block} have the same contraction pattern, differing only by an interchange of the time arguments of the two Green's functions. The same relation holds between the second and fourth terms. It is therefore sufficient to analyze only the first two terms.

For the first term, rather than placing the atom pair $\expval{IJ}$ in the outermost loop, it is more advantageous to place $\expval{IL}$ there, which enables the contractions over the indices $K$ and $J$ to be carried out separately.
For the second term, the loops over $K$ and $L$ can likewise be decoupled by exploiting the symmetric contraction structure of $\mathbf{C}^{\cMu}_{IK}G_{JK}$ and $\mathbf{C}^{\cNu}_{JL}G_{IL}$.
This reorganization removes one level of nested looping and thus improves upon the algorithm proposed in Ref.~\citenum{ShiR24} (Algorithm~2 therein).
The resulting implementation of Eq.~\eqref{eq:chi0-mn-ap-block} is summarized in Algorithm~\ref{alg:chi-1-all-terms} and has been implemented in LibRI,\cite{libri-repo} an open-source package for efficient tensor contractions in advanced electronic-structure methods.

\begin{algorithm}
\caption{\label{alg:chi-1-all-terms}
  Algorithm for evaluating Eq.~\eqref{eq:chi0-mn-ap-block},
  i.e., the real-space imaginary-time response function $\chi^0(\ii\tau)$ in the auxiliary-basis representation at a given imaginary-time point.
  Here, $\mathcal{N}\!\left[I(\bR)\right]$ denotes the set of neighboring atoms of atom $I$ located in the unit cell $\bR$.
}
\ForEach{$\expval{\cMu(\mathbf{0}), I=\cMu}$}{
  \ForEach{$L(\bR_2)$}{
    Initialize temporaries $\mathbf{X}^{\cMu-}_{IL}$, $\mathbf{X}^{\cMu+}_{IL}$\;
    \ForEach{$K(\bR_1)\in\mathcal{N}\left[\cMu(\mathbf{0})\right]$}{
      \If{$\max{\abs{\mathbf{G}^{-}_{LK}(\bR_1-\bR_2)}} > \eta_G$}{
        $\mathbf{X}^{\cMu-}_{IL}(\bR_2) \mathrel{+}= \mathbf{C}^{\cMu(\mathbf{0})}_{I(\mathbf{0}),K(\bR_1)} \mathbf{G}^{-}_{LK}(\bR_1-\bR_2)$\;
      }
      \If{$\max{\abs{\mathbf{G}^{+}_{KL}(\bR_2-\bR_1)}} > \eta_G$}{
        $\mathbf{X}^{\cMu+}_{IL}(\bR_2) \mathrel{+}= \mathbf{C}^{\cMu(\mathbf{0})}_{I(\mathbf{0}),K(\bR_1)} \mathbf{G}^{+}_{KL}(\bR_2-\bR_1)$\;
      }
    }
    \ForEach{$\expval{\cNu(\bR)\in\mathcal{N}\left[L(\bR_2)\right], J=\cNu}$}{
      \If{$\max{\abs{\mathbf{G}^{+}_{IJ}(\bR)}} > \eta_G$}{
        $\boldsymbol{\chi}^0_{\cMu\cNu}(\bR) \mathrel{+}= \mathbf{G}^{+}_{IJ}(\bR)\,
        \mathbf{X}^{\cMu-}_{IL}(\bR_2)\,
        \mathbf{C}^{\cNu(\bR)}_{J(\bR),L(\bR_2)}$
        \tcp*{$\left[ \text{c.f. 1st term} \right]$}
      }
      \If{$\max{\abs{\mathbf{G}^{-}_{JI}(-\bR)}} > \eta_G$}{
        $\boldsymbol{\chi}^0_{\cMu\cNu}(\bR) \mathrel{+}= \mathbf{G}^{-}_{JI}(-\bR)\,
        \mathbf{X}^{\cMu+}_{IL}(\bR_2)\,
        \mathbf{C}^{\cNu(\bR)}_{J(\bR),L(\bR_2)}$
        \tcp*{$\left[ \text{c.f. 3rd term} \right]$}
      }
    }
  }

  \ForEach{$\expval{\cNu(\bR), J=\cNu}$}{
    Initialize temporaries
    $\mathbf{Y}^{\cNu-}_{IJ}$, $\mathbf{Y}^{\cNu+}_{IJ}$,
    $\mathbf{Q}^{\cMu-}_{IJ}$, $\mathbf{Q}^{\cMu+}_{IJ}$\;
    \ForEach{$L(\bR_2)\in\mathcal{N}\left[\cNu(\bR)\right]$}{
      \If{$\max{\abs{\mathbf{G}^{+}_{IL}(\bR_2)}} > \eta_G$}{
        $\mathbf{Y}^{\cNu+}_{IJ}(\bR) \mathrel{+}= \mathbf{C}^{\cNu(\bR)}_{J(\bR),L(\bR_2)} \mathbf{G}^{+}_{IL}(\bR_2)$\;
      }
      \If{$\max{\abs{\mathbf{G}^{-}_{LI}(-\bR_2)}} > \eta_G$}{
        $\mathbf{Y}^{\cNu-}_{IJ}(\bR) \mathrel{+}= \mathbf{C}^{\cNu(\bR)}_{J(\bR),L(\bR_2)} \mathbf{G}^{-}_{LI}(-\bR_2)$\;
      }
    }
    \ForEach{$K(\bR_1)\in\mathcal{N}\left[\cMu(\mathbf{0})\right]$}{
      \If{$\max{\abs{\mathbf{G}^{-}_{JK}(\bR_1-\bR)}} > \eta_G$}{
        $\mathbf{Q}^{\cMu-}_{IJ}(\bR) \mathrel{+}= \mathbf{C}^{\cMu(\mathbf{0})}_{I(\mathbf{0}),K(\bR_1)} \mathbf{G}^{-}_{JK}(\bR_1-\bR)$\;
      }
      \If{$\max{\abs{\mathbf{G}^{+}_{KJ}(\bR-\bR_1)}} > \eta_G$}{
        $\mathbf{Q}^{\cMu+}_{IJ}(\bR) \mathrel{+}= \mathbf{C}^{\cMu(\mathbf{0})}_{I(\mathbf{0}),K(\bR_1)} \mathbf{G}^{+}_{KJ}(\bR-\bR_1)$\;
      }
    }
    $\boldsymbol{\chi}^0_{\cMu\cNu}(\bR) \mathrel{+}=
    \mathbf{Q}^{\cMu-}_{IJ}(\bR)\mathbf{Y}^{\cNu+}_{IJ}(\bR)$
    \tcp*{$\left[ \text{c.f. 2nd term} \right]$}
    $\boldsymbol{\chi}^0_{\cMu\cNu}(\bR) \mathrel{+}=
    \mathbf{Q}^{\cMu+}_{IJ}(\bR)\mathbf{Y}^{\cNu-}_{IJ}(\bR)$
    \tcp*{$\left[ \text{c.f. 4th term} \right]$}
  }
}
\end{algorithm}

\subsubsection{Self-energy}\label{sssec:contraction-sigc}
We now turn to the computation of the self-energy.
As noted above, the exchange self-energy and the imaginary-time correlation self-energy have the same algebraic structure. We therefore focus here on the correlation part only.

\begin{figure}
\includegraphics[width=0.96\linewidth]{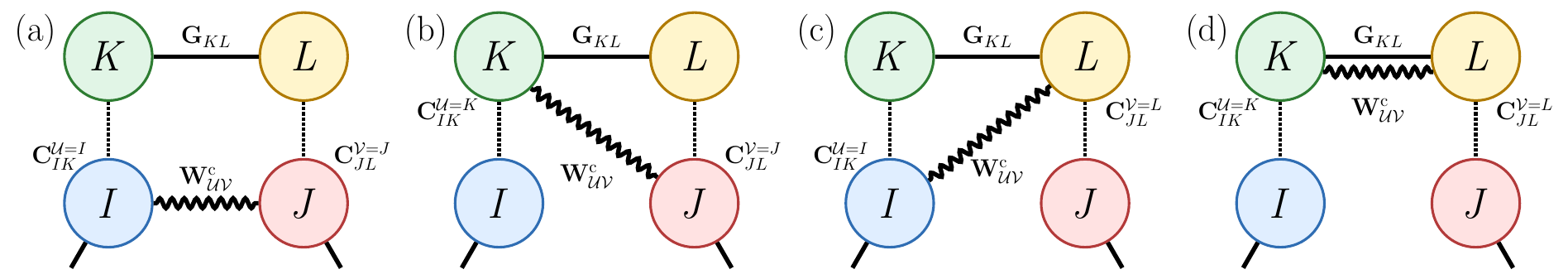}
\caption{\label{fig:sigmac-IJ-pattern}
  Schematic diagrams of the contributions from an atom quartet $\expval{IJKL}$ to the atom-pair block of the correlation self-energy $\boldsymbol{\Sigma}^{\mathrm{c}}_{IJ}$ in Eq.~\eqref{eq:sigmac-ij-Rt-label-sum}.
  From left to right, the diagrams correspond to Eqs.~\eqref{eq:sigmac-ij-Rt-label-IJ}, \eqref{eq:sigmac-ij-Rt-label-KJ}, \eqref{eq:sigmac-ij-Rt-label-IL}, and \eqref{eq:sigmac-ij-Rt-label-KL}, respectively.
  The line conventions are the same as those in Fig.~\ref{fig:chi0-tensor-contract-pattern}.
  The imaginary-time arguments of $\mathbf{G}$ and $\mathbf{W}^{\mathrm{c}}$ are omitted for brevity.
}
\end{figure}

To expose the contraction structure and facilitate an efficient parallel implementation, we rewrite the correlation self-energy, analogous to $\boldsymbol{\chi}^0_{\cMu\cNu}$, in terms of atom-pair-block contributions as
\begin{equation}\label{eq:sigmac-ij-Rt-label-sum}
\begin{aligned}
\boldsymbol{\Sigma}^{\mathrm{c}}_{IJ} = \sum_{KL} \left[\boldsymbol{\Sigma}^{\mathrm{c},\mathrm{(A)}}_{IJKL} +
  \boldsymbol{\Sigma}^{\mathrm{c},\mathrm{(B)}}_{IJKL} +
  \boldsymbol{\Sigma}^{\mathrm{c},\mathrm{(C)}}_{IJKL} +
  \boldsymbol{\Sigma}^{\mathrm{c},\mathrm{(D)}}_{IJKL}
\right],
\end{aligned}
\end{equation}
where, for simplicity, the common time argument is omitted.
The four terms are defined as
\begin{subequations}\label{eq:sigmac-ij-Rt-labels}
\begin{equation}\label{eq:sigmac-ij-Rt-label-IJ}
\boldsymbol{\Sigma}^{\mathrm{c},\mathrm{(A)}}_{IJKL} \equiv \boldsymbol{\Sigma}^{\mathrm{c}}_{\underline{IJ};KL} = - \mathbf{G}_{KL} \mathbf{C}^{\cMu=I}_{IK} \bWc_{\cMu=I,\cNu=J} \mathbf{C}^{\cNu=J}_{JL}
\end{equation}
\begin{equation}\label{eq:sigmac-ij-Rt-label-KJ}
\boldsymbol{\Sigma}^{\mathrm{c},\mathrm{(B)}}_{IJKL} \equiv
\boldsymbol{\Sigma}^{\mathrm{c}}_{I\underline{J};\underline{K}L} = - \mathbf{G}_{KL} \mathbf{C}^{\cMu=K}_{IK} \bWc_{\cMu=K,\cNu=J} \mathbf{C}^{\cNu=J}_{JL} 
\end{equation}
\begin{equation}\label{eq:sigmac-ij-Rt-label-IL}
\boldsymbol{\Sigma}^{\mathrm{c},\mathrm{(C)}}_{IJKL} \equiv
\boldsymbol{\Sigma}^{\mathrm{c}}_{\underline{I}J;K\underline{L}} = - \mathbf{G}_{KL} \mathbf{C}^{\cMu=I}_{IK} \bWc_{\cMu=I,\cNu=L} \mathbf{C}^{\cNu=L}_{JL}
\end{equation}
\begin{equation}\label{eq:sigmac-ij-Rt-label-KL}
\boldsymbol{\Sigma}^{\mathrm{c},\mathrm{(D)}}_{IJKL} \equiv
\boldsymbol{\Sigma}^{\mathrm{c}}_{IJ;\underline{KL}} = - \mathbf{G}_{KL} \mathbf{C}^{\cMu=K}_{IK} \bWc_{\cMu=K,\cNu=L} \mathbf{C}^{\cNu=L}_{JL}
\end{equation}
\end{subequations}
In this notation, the underlined atom labels indicate the atoms on which the auxiliary-basis-function indices reside within a given atom quartet $\expval{IJKL}$.
The four contributions from each quartet to the correlation self-energy block $\boldsymbol{\Sigma}^{\mathrm{c}}_{IJ}$ are illustrated in Fig.~\ref{fig:sigmac-IJ-pattern}.

In this form, a straightforward implementation would distribute and loop over the atom pairs $\expval{IJ}$ associated with the uncontracted one-electron NAO indices, and for each quartet traverse the four screened Coulomb blocks
$\{\bWc_{\cMu\cNu}\mid \cMu\in\{I,K\},\,\cNu\in\{J,L\}\}$
in the innermost loop.
However, this loop structure is unfavorable for parallelization because it leads to unnecessary replication of $\bWc_{\cMu\cNu}$ across multiple processes.
Because the screened Coulomb interaction is stored in auxiliary-basis atom-pair blocks, these blocks are typically much larger than the corresponding one-electron NAO blocks, so their unnecessary replication can result in significant memory overhead.
A similar issue was pointed out in Ref.~\citenum{LinP21JCTC} in the context of computing the nonlocal exchange operator.

\begin{figure}
\includegraphics[width=0.96\linewidth]{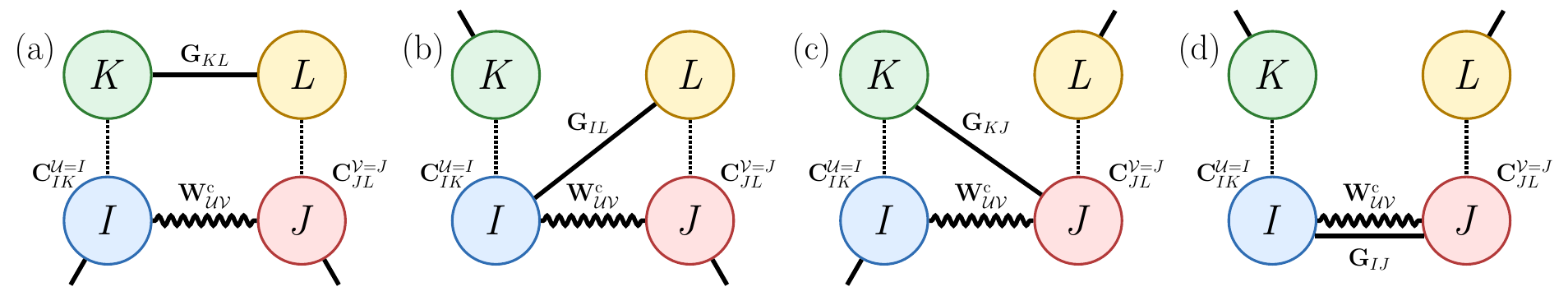}
\caption{\label{fig:sigmac-rearrange-pattern}
  Schematic diagrams of the rearranged terms in Eq.~\eqref{eq:sigmac-rearrange-labels}, illustrating how a fixed screened-interaction block $\bWc_{\cMu=I,\cNu=J}$ contributes to multiple atom-pair blocks of the correlation self-energy from an atom quartet $\expval{IJKL}$.
  The line conventions are the same as those in Fig.~\ref{fig:sigmac-IJ-pattern}, and the imaginary-time arguments of $\mathbf{G}$ and $\mathbf{W}^{\mathrm{c}}$ are omitted for brevity.
}
\end{figure}

A more efficient strategy is to distribute and loop over the atom pairs $\expval{\cMu\cNu}$ on which the ABF indices reside and, in the innermost loop, accumulate the contributions from $\bWc_{\cMu\cNu}$ to the set of self-energy blocks
$\{\boldsymbol{\Sigma}^{\mathrm{c}}_{AB} \mid A\in\{I,K\},\, B\in\{J,L\}\}$.
For a given pair $\expval{\cMu\cNu}$, the LRI approximation constrains two atoms in each atom quartet $\expval{IJKL}$.
In the cases $I=\cMu$ and $J=\cNu$, the relevant contributions are
\begin{equation}\label{eq:sigmac-rearrange-labels}
\begin{aligned}
\boldsymbol{\Sigma}^{\mathrm{c}}_{IJ} \;\leftarrow\; & \boldsymbol{\Sigma}^{\mathrm{c},\mathrm{(A)}}_{IJKL}
= - \mathbf{G}_{KL} \mathbf{C}^{\cMu=I}_{IK} \bWc_{\cMu=I,\cNu=J} \mathbf{C}^{\cNu=J}_{JL} \\
\boldsymbol{\Sigma}^{\mathrm{c}}_{KJ} \;\leftarrow\; & \boldsymbol{\Sigma}^{\mathrm{c},\mathrm{(B)}}_{KJIL}
= - \mathbf{G}_{IL} \mathbf{C}^{\cMu=I}_{IK} \bWc_{\cMu=I,\cNu=J} \mathbf{C}^{\cNu=J}_{JL} \\
\boldsymbol{\Sigma}^{\mathrm{c}}_{IL} \;\leftarrow\; & \boldsymbol{\Sigma}^{\mathrm{c},\mathrm{(C)}}_{ILKJ}
= - \mathbf{G}_{KJ} \mathbf{C}^{\cMu=I}_{IK} \bWc_{\cMu=I,\cNu=J} \mathbf{C}^{\cNu=J}_{JL} \\
\boldsymbol{\Sigma}^{\mathrm{c}}_{KL} \;\leftarrow\; & \boldsymbol{\Sigma}^{\mathrm{c},\mathrm{(D)}}_{KLIJ}
= - \mathbf{G}_{IJ} \mathbf{C}^{\cMu=I}_{IK} \bWc_{\cMu=I,\cNu=J} \mathbf{C}^{\cNu=J}_{JL}
\end{aligned}
\end{equation}
The equivalence of Eq.~\eqref{eq:sigmac-rearrange-labels} to Eq.~\eqref{eq:sigmac-ij-Rt-label-sum} for constructing the full self-energy matrix $\boldsymbol{\Sigma}^{\mathrm{c}}$ can be seen by permuting $I\leftrightarrow K$ and/or $J\leftrightarrow L$ in Eq.~\eqref{eq:sigmac-rearrange-labels}.
This generates a total of sixteen contributions to
$\{\boldsymbol{\Sigma}^{\mathrm{c}}_{AB} \mid A\in\{I,K\},\, B\in\{J,L\}\}$,
among which four contribute to $\boldsymbol{\Sigma}^{\mathrm{c}}_{IJ}$ and are identical to the terms in Eq.~\eqref{eq:sigmac-ij-Rt-label-sum}.
Importantly, the four terms in Eq.~\eqref{eq:sigmac-rearrange-labels} differ only in the Green's-function block and share the same $\mathbf{C}\bWc\mathbf{C}$ factor.
This reorganization enables efficient parallelization over atom quartets and avoids redundant duplication of screened Coulomb blocks.
The rearranged scheme is illustrated in Fig.~\ref{fig:sigmac-rearrange-pattern}.

An efficient implementation of Eq.~\eqref{eq:sigmac-rearrange-labels} requires treating the four terms separately because their contraction patterns are different.
The pattern of $\boldsymbol{\Sigma}^{\mathrm{c},\mathrm{(A)}}$ [Fig.~\ref{fig:sigmac-rearrange-pattern}(a)] is the same as that of the first term in $\boldsymbol{\chi}^0$ [Fig.~\ref{fig:chi0-tensor-contract-pattern}(a)], except that here the NAO indices associated with atoms $I$ and $J$ remain open, whereas the ABF indices are contracted.
Therefore, the same three-layer loop structure can be used for $\boldsymbol{\Sigma}^{\mathrm{c},\mathrm{(A)}}$.

For $\boldsymbol{\Sigma}^{\mathrm{c},\mathrm{(B)}}$ [Fig.~\ref{fig:sigmac-rearrange-pattern}(b)], it is advantageous to place the loop over $\expval{IJ}$ in the outermost layer, so that the sparsity of $\bWc_{\cMu=I,\cNu=J}$ can be checked at an early stage.
In the inner loop, the atom index $L$ is contracted first by forming the intermediate array
$\mathbf{S}^{\cNu}_{IJ} = \sum_L \mathbf{G}_{IL}\mathbf{C}^{\cNu}_{JL}$.
This is followed by the second inner loop over $K$, which completes the evaluation of the term.
The term $\boldsymbol{\Sigma}^{\mathrm{c},\mathrm{(C)}}$ [Fig.~\ref{fig:sigmac-rearrange-pattern}(c)] can be evaluated in an analogous manner because of its symmetric contraction pattern.

Finally, for $\boldsymbol{\Sigma}^{\mathrm{c},\mathrm{(D)}}$ [Fig.~\ref{fig:sigmac-rearrange-pattern}(d)], an efficient three-layer loop structure can also be constructed.
Its contraction pattern is diagrammatically symmetric to that of $\boldsymbol{\Sigma}^{\mathrm{c},\mathrm{(A)}}$ [Fig.~\ref{fig:sigmac-rearrange-pattern}(a)] in terms of one-electron NAO indices, and the corresponding loop hierarchy can be obtained from that of $\boldsymbol{\Sigma}^{\mathrm{c},\mathrm{(A)}}$ by interchanging $I \leftrightarrow K$ and $J \leftrightarrow L$.
However, the contraction with $\bWc$ must be carried out when the inner loop reaches the second ABF index associated with $\bWc$, which is $I$ for $\boldsymbol{\Sigma}^{\mathrm{c},\mathrm{(D)}}$.
Accordingly, whereas for $\boldsymbol{\Sigma}^{\mathrm{c},\mathrm{(A)}}$ one first forms an intermediate of the form $\mathbf{C}\mathbf{G}$ and then evaluates
$\boldsymbol{\Sigma}^{\mathrm{c}} \leftarrow -(\mathbf{C}\mathbf{G})\bWc\mathbf{C}$,
for $\boldsymbol{\Sigma}^{\mathrm{c},\mathrm{(D)}}$ one first constructs an intermediate of the form $\mathbf{C}\mathbf{G}\bWc$ and then performs the final contraction with $\mathbf{C}$.

In addition to the contraction patterns discussed above, the sparsity of $\mathbf{G}$ and $\bWc$ can be exploited to skip evaluations involving atom-pair blocks whose elements are negligibly small, thereby further reducing the computational cost.
The corresponding procedures for $\boldsymbol{\Sigma}^{\mathrm{c},\mathrm{(A)}}$, $\boldsymbol{\Sigma}^{\mathrm{c},\mathrm{(B)}}$, $\boldsymbol{\Sigma}^{\mathrm{c},\mathrm{(C)}}$, and $\boldsymbol{\Sigma}^{\mathrm{c},\mathrm{(D)}}$ are summarized in Algorithms~\ref{alg:sigmac-rearrange-IJ}, \ref{alg:sigmac-rearrange-KJ}, \ref{alg:sigmac-rearrange-IL}, and \ref{alg:sigmac-rearrange-KL}, respectively.

For the exchange self-energy $\Sigma^{\mathrm{x}}$, i.e., the nonlocal exchange operator, the algorithm is almost identical to that for $\Sigma^{\mathrm{c}}$.
The difference is that the density matrix $\mathbf{D}$ and the bare Coulomb matrix $\mathbf{V}$ enter in place of the Green's function and the screened Coulomb interaction, respectively.
Accordingly, the computational cost of $\Sigma^{\mathrm{x}}$ is comparable to that of $\Sigma^{\mathrm{c}}$ at a single imaginary-time point.
In practice, however, $\Sigma^{\mathrm{c}}$ must be evaluated at multiple points on the imaginary-time axis for both positive and negative times.
Therefore, evaluating $\Sigma^{\mathrm{c}}$ over all imaginary-time points costs roughly $2N_{\tau}$ times as much as evaluating $\Sigma^{\mathrm{x}}$.
This factor amounts to 32 and 64 for 16-point and 32-point minimax grids, respectively.
Thus, the total self-energy calculation is dominated by the correlation part.

\begin{algorithm}
\caption{\label{alg:sigmac-rearrange-IJ}
Algorithm for evaluating the contribution $\boldsymbol{\Sigma}^{\mathrm{c},\mathrm{(A)}}$ to the correlation self-energy in Eq.~\eqref{eq:sigmac-rearrange-labels}.}
\ForEach{$I(\mathbf{0})$}{
  Constrain $\cMu = I$\;
  \ForEach{$L(\bR_2)$}{
    Initialize temporary $\mathbf{A}^{\cMu}_{IL}$\;
    \ForEach{$K(\bR_1) \in \mathcal{N}\left[I(\mathbf{0})\right]$}{
      \If{$\max{\abs{\mathbf{G}_{KL}(\bR_2-\bR_1)}} > \eta_G$}{
        $\mathbf{A}^{\cMu}_{IL}(\bR_2) \mathrel{+}= \mathbf{C}^{\cMu(\mathbf{0})}_{I(\mathbf{0}),K(\bR_1)} \mathbf{G}_{KL}(\bR_2-\bR_1)$\;
      }
    }
    \ForEach{$J(\bR) \in \mathcal{N}\left[L(\bR_2)\right]$}{
      Constrain $\cNu = J$\;
      \If{$\max{\abs{\bWc_{\cMu\cNu}(\bR)}} > \eta_W$}{
        $\boldsymbol{\Sigma}^{\mathrm{c}}_{IJ}(\bR) \mathrel{+}= -\bWc_{\cMu\cNu}(\bR)\,
        \mathbf{A}^{\cMu}_{IL}(\bR_2)\,
        \mathbf{C}^{\cNu(\bR)}_{J(\bR),L(\bR_2)}$\;
      }
    }
  }
}
\end{algorithm}

\begin{algorithm}
\caption{\label{alg:sigmac-rearrange-KJ}
Algorithm for evaluating the contribution $\boldsymbol{\Sigma}^{\mathrm{c},\mathrm{(B)}}$ to the correlation self-energy in Eq.~\eqref{eq:sigmac-rearrange-labels}.
}
\ForEach{$J(\mathbf{\bR})$}{
  Constrain $\cNu = J$\;
  \ForEach{$I(\mathbf{0})$}{
    Constrain $\cMu = I$, initialize temporary $\mathbf{S}^{\cNu}_{IJ}$\;
    \If{$\max{\abs{\bWc_{\cMu\cNu}(\bR)}} < \eta_W$}{
      \Continue\;
    }
    \ForEach{$L(\bR_2) \in \mathcal{N}\left[J(\bR)\right]$}{
      \If{$\max{\abs{\mathbf{G}_{IL}(\bR_2)}} > \eta_G$}{
        $\mathbf{S}^{\cNu}_{IJ}(\bR) \mathrel{+}= \mathbf{G}_{IL}(\bR_2)\,
        \mathbf{C}^{\cNu(\bR)}_{J(\bR),L(\bR_2)}$\;
      }
    }
    \ForEach{$K(\bR_1) \in \mathcal{N}\left[I(\mathbf{0})\right]$}{
      $\boldsymbol{\Sigma}^{\mathrm{c}}_{KJ}(\bR-\bR_1) \mathrel{+}=
      -\mathbf{C}^{\cMu(\mathbf{0})}_{I(\mathbf{0}),K(\bR_1)}
      \bWc_{\cMu\cNu}(\bR)\,
      \mathbf{S}^{\cNu}_{IJ}(\bR)$\;
    }
  }
}
\end{algorithm}

\begin{algorithm}
\caption{\label{alg:sigmac-rearrange-IL}
Algorithm for evaluating the contribution $\boldsymbol{\Sigma}^{\mathrm{c},\mathrm{(C)}}$ to the correlation self-energy in Eq.~\eqref{eq:sigmac-rearrange-labels}.
}
\ForEach{$I(\mathbf{0})$}{
  Constrain $\cMu = I$\;
  \ForEach{$J(\mathbf{\bR})$}{
    Constrain $\cNu = J$, initialize temporary $\mathbf{T}^{\cMu}_{IJ}$\;
    \If{$\max{\abs{\bWc_{\cMu\cNu}(\bR)}} < \eta_W$}{
      \Continue\;
    }
    \ForEach{$K(\bR_1) \in \mathcal{N}\left[I(\mathbf{0})\right]$}{
      \If{$\max{\abs{\mathbf{G}_{KJ}(\bR-\bR_1)}} > \eta_G$}{
        $\mathbf{T}^{\cMu}_{IJ}(\bR) \mathrel{+}= \mathbf{C}^{\cMu(\mathbf{0})}_{I(\mathbf{0}),K(\bR_1)}
        \mathbf{G}_{KJ}(\bR-\bR_1)$\;
      }
    }
    \ForEach{$L(\bR_2) \in \mathcal{N}\left[J(\bR)\right]$}{
      $\boldsymbol{\Sigma}^{\mathrm{c}}_{IL}(\bR_2) \mathrel{+}=
      -\mathbf{T}^{\cMu}_{IJ}(\bR)\,
      \bWc_{\cMu\cNu}(\bR)\,
      \mathbf{C}^{\cNu(\bR)}_{J(\bR),L(\bR_2)}$\;
    }
  }
}
\end{algorithm}

\begin{algorithm}
\caption{\label{alg:sigmac-rearrange-KL}
Algorithm for evaluating the contribution $\boldsymbol{\Sigma}^{\mathrm{c},\mathrm{(D)}}$ to the correlation self-energy in Eq.~\eqref{eq:sigmac-rearrange-labels}.
}
\ForEach{$K(\bR_1)$}{
  \ForEach{$J(\mathbf{\bR})$}{
    Constrain $\cNu = J$, initialize temporary $\mathbf{B}^{\cNu}_{KJ}$\;
    \ForEach{$I(\mathbf{0}) \in \mathcal{N}\left[K(\bR_1)\right]$}{
      Constrain $\cMu = I$\;
      \If{$\max{\abs{\mathbf{G}_{IJ}(\bR)}} > \eta_G$ \textnormal{and}
          $\max{\abs{\bWc_{\cMu\cNu}(\bR)}} > \eta_W$}{
        $\mathbf{B}^{\cNu}_{KJ}(\bR-\bR_1) \mathrel{+}=
        \mathbf{C}^{\cMu(\mathbf{0})}_{I(\mathbf{0}),K(\bR_1)}
        \bWc_{\cMu\cNu}(\bR)\,
        \mathbf{G}_{IJ}(\bR)$\;
      }
    }
    \ForEach{$L(\bR_2) \in \mathcal{N}\left[J(\bR)\right]$}{
      $\boldsymbol{\Sigma}^{\mathrm{c}}_{KL}(\bR_2-\bR_1) \mathrel{+}=
      -\mathbf{B}^{\cNu}_{KJ}(\bR-\bR_1)
      \mathbf{C}^{\cNu(\bR)}_{J(\bR),L(\bR_2)}$\;
    }
  }
}
\end{algorithm}

The adopted three-layer loop structure for evaluating the self-energy $\Sigma$ is an improved version of the algorithm for $\Sigma^{\mathrm{x}}$ proposed in Ref.~\citenum{LinP21JCTC}.
The resulting self-energy algorithms have also been implemented in LibRI.\cite{libri-repo}

\subsection{Treatment of $\Gamma$-point singularity of the Coulomb matrix}\label{ssec:finite-size}
The bare Coulomb matrix $V_{\mu\nu}(\bq)$ [Eq.~\eqref{eq:v-mn-q}] has diverging elements as $\bq\to 0$ due to the long-range nature of the electrostatic interaction, and careful treatment of the $\Gamma$ point singularity is required when dealing with periodic systems.
\JCTC{%
More explicitly, in three-dimensional systems, the leading term of the general lattice sum for \(V_{\mu\nu}(\bq)\) scales as \(|\bq|^{L-2}\) (see, e.g., the second term of Eq.~(44) in Ref.~\citenum{NijboerB57}), where \(L=l(\mu)+l(\nu)\) and \(l(\mu)\) is the angular momentum associated with the ABF \(P_{\mu}\).
This asymptotic behavior was also discussed for localized muffin-tin basis functions in the LAPW framework in Ref.~\citenum{FriedrichC09}.
A more comprehensive discussion on the convergence of the lattice sum is provided in Ref.~\citenum{StolarczykLZ82} and Appendix D of Ref.~\citenum{PasquierR25}.
Consequently, the Coulomb matrix in the ABF representation contains divergent elements when \(L<2\), i.e., when both orbitals are \(s\)-type (\(|\bq|^{-2}\)) or when one orbital is \(s\)-type and the other is \(p\)-type (\(|\bq|^{-1}\)).
Although this singularity is canceled by the \(|\bq|^2\) behavior of the response function \(\chi(\bq\to0)\) when forming the symmetrized dielectric matrix in Eq.~\eqref{eq:symeps}, so that the dielectric matrix is formally non-singular, numerical precision may still be compromised if the \(\bq=0\) singularity is not handled properly.
}%
A minimalist approach is to exclude the $\Gamma$ point completely from the BZ sampling, which circumvents the necessity of special treatment for singular matrices.
However, this results in a slow $\mathcal{O}(\Nk^{-1/3})$ convergence in the self-energy as the number of k-points $\Nk$ increases, and brings considerable finite-size errors when a moderate $\bk$-point mesh of correlation calculations, e.g., $8\times8\times8$ for three-dimensional bulk systems, is used.

To handle the $\Gamma$-point singularity and speed up the convergence with respect to the $\bk$-point mesh,
we follow the methodology discussed in Ref. \citenum{RenX21}, which is recapitulated below.
First, the bare Coulomb matrices are constructed with the singularity lifted \JCTC{based on an auxiliary-function approach}.\cite{BroqvistP09,Levchenko/etal:2015}
This ensures that the regularized form of the Coulomb matrix, denoted by $\bar{V}_{\mu\nu}(\bq)$, is finite at $\bq\to0$ so that the matrix 
can be diagonalized at $\bq=0$.
We explicitly solve the eigenvalue problem of the Hermitian matrix $\bar{V}_{\mu\nu}(\bq=0)$, namely
\begin{equation}\label{eq:v-diagonalize}
\sum_{\nu} \bar{V}_{\mu\nu}(\bq=0) X_{\nu\lambda} = v_{\lambda} X_{\mu\lambda}
\end{equation}
where $v_{\lambda}$ is the real-valued eigenvalue of $\bar{V}_{\mu\nu}(\bq=0)$
associated with the eigenvector $X_{\mu\lambda}, \mu=1,\ldots,N_{\mathrm{aux}}$.
The diagonalized Coulomb matrix $\bar{V}^{v}_{\lambda\lambda'} = v_{\lambda}\delta_{\lambda\lambda'}$ can be connected to the plane-wave representation, i.e., $V_{\bG\bG'}(\bq\to0) = 4\pi\delta_{\bG\bG'}/\abs{\bG}^2$, where the singularity occurs solely at the head element $V_{\bG=0,\bG'=0}$.
This motivates a unitary transformation of the symmetrized dielectric matrix $\tilde{\varepsilon}_{\mu\nu}$ to its Coulomb-eigenvector representation $\tilde{\varepsilon}^{v}_{\lambda\lambda'}$ via
\begin{equation}
\tilde{\varepsilon}^{v}_{\lambda\lambda'} = \sum_{\mu\nu} X^\ast
_{\mu\lambda}  \tilde{\varepsilon}_{\mu\nu}(\bq=0) X_{\nu\lambda'}.
\end{equation}
With $v_{\lambda}, \lambda=1,\ldots,N_{\mathrm{aux}}$ sorted in descending order,
we approximate the eigenvector $X_{\mu1}$ as the $\bG=0$ plane-wave.
This allows us to apply corrections devised for $\tilde{\varepsilon}_{\bG\bG'}(\bq\to0)$ analogously to $\tilde{\varepsilon}^{v}_{\lambda\lambda'}$.
In this work, we apply the head term correction in the isotropic approximation
\begin{equation}\label{eq:head-term-iso}
\tilde{\varepsilon}^{v}_{\lambda=1,\lambda'=1}(\ii\omega) = \frac{1}{3}\left[H^{11}(\ii\omega) + H^{22}(\ii\omega) + H^{33}(\ii\omega)\right]
\end{equation}
with
\begin{equation}\label{eq:head-tensor-kp}
H^{\alpha\beta}(\ii\omega) = \delta_{\alpha\beta} - 2\frac{4\pi}{N_{\bk}\Omega} 
\sum_{\bk} \left\lbrace \sum_{n}
\frac{f'_{n\bk} p^{\bk *}_{\alpha, nn} p^{\bk}_{\beta, nn}}{\omega^2} +
\sum_{m<n}
\frac{2(f_{m\bk} - f_{n\bk})p^{\bk *}_{\alpha, mn} p^{\bk}_{\beta, mn}}
{\left[\left(\epsilon_{m\bk} - \epsilon_{n\bk}\right)^2+\omega^2\right](\epsilon_{m\bk} - \epsilon_{n\bk})}
\right\rbrace
\end{equation}
where $\Omega$ is the volume of unit cell,
$f'_{n\bk} = \left.\partial f / \partial \epsilon\right|_{\epsilon_{n\bk}}$ the energy derivative of the occupation number of state $\ket{n\bk}$,
and
\begin{equation}\label{eq:mommat}
p^{\bk}_{\alpha, nm}
= \sum_{ij} c^{i*}_{n\bk} c^{j}_{m\bk} \sum_{\bR}\ee^{\ii\bk\cdot\bR} \mel{\varphi_{i\mathbf{0}}}{\nabla_{\alpha}}{\varphi_{j\bR}}
\end{equation}
is the momentum matrix element in Kohn-Sham representation. The first term in the curly brackets vanishes for insulating systems.
Eq.~\eqref{eq:head-tensor-kp} is in fact the macroscopic dielectric tensor without taking into account the local field effect.
In principle, the correction should also be applied to the wing terms, namely $\tilde{\varepsilon}^{v}_{\lambda=1,\lambda'>1}$ and $\tilde{\varepsilon}^{v}_{\lambda>1,\lambda'=1}$.
For the cubic systems we investigate here, the effect on band gaps due to the wing term correction is negligible. The isotropic approximation also works well for these systems.
After the correction is applied, $\tilde{\varepsilon}^{v}_{\lambda\lambda'}$ is rotated back to its ABF representation $\tilde{\varepsilon}_{\mu\nu}$ for further computation.

The long-range nature of bare Coulomb matrices also affects the calculation of self-energy. In the canonical $k$-space implementation, this amounts to the singularity in $\Wc_{\mu\nu}(\bq=0)$ constructed from $V_{\mu\nu}(\bq=0)$ in Eq.~\eqref{eq:w0-mu-qw-symeps}.
In the current real-space approach, it results in a non-convergent summation over the cell vectors in Eq.~\eqref{eq:vexx-R} and Eq.~\eqref{eq:sigmac-ij-Rt}.
To circumvent this issue, we follow Ref. \citenum{RenX21} and adopt the spherically truncated Coulomb kernel for $V_{\mu\nu}$ in Eq.~\eqref{eq:vexx-R} and Eq.~\eqref{eq:w0-mu-qw-symeps}.
Particularly, instead of an abrupt truncation in real space as originally proposed by Spencer and Alavi \cite{SpencerJ08}, an attenuated Coulomb kernel
\begin{equation}
v^{c}(r) = \frac{\erfc(\gamma r)}{r} + \frac{\erf(\gamma r)}{2}\erfc\left[\frac{\ln(r)-\ln(R_{\mathrm{cut}})}{\ln(R_{\mathrm{w}})}\right]
\end{equation}
is utilized to replace the bare one in Eq.~\eqref{eq:v-mn-q} to improve the numerical stability of real-space grid integration.
The screening parameters are taken as the same values as in Ref. \citenum{RenX21},
i.e. $\gamma=5.0 / R_{\mathrm{cut}}\text{ Bohr}^{-1}$  and $R_w=1.092\text{ Bohr}$.

\subsection{Iterative solution of the quasi-particle equation}

After all required components have been computed, the quasi-particle equation, Eq.~\eqref{eq:quasi-particle-equation} can be explicitly written as
\begin{equation}\label{eq:quasi-particle-equation-decomp}
\omega = \epsilon_{n\bk} - v^{\xc}_{n\bk} + \Sigma^{\mathrm{x}}_{nn}(\bk) + \Sigma^{\mathrm{c}}_{nn}(\bk,\omega)
\end{equation}
where
\begin{equation}
\Sigma^{\mathrm{x}}_{nn}(\bk) = \sum_{ij} c^{i*}_{n\bk} c^{j}_{n\bk} \sum_{\bR} \ee^{\ii\bk\cdot\bR} \Sigma^{\mathrm{x}}_{ij}(\bR)
\end{equation}
with $\Sigma^{\mathrm{x}}_{ij}(\bR)$ obtained from Eq.~\eqref{eq:vexx-R},
and $\Sigma^{\mathrm{c}}_{nn}(\bk,\omega)$ obtained by analytic continuation (Sec.~\ref{sssec:ac}) of
\begin{equation}
\Sigma^{\mathrm{c}}_{nn}(\bk,\ii\omega) = \sum_{ij} c^{i*}_{n\bk} c^{j}_{n\bk} \sum_{\bR} \ee^{\ii\bk\cdot\bR} \Sigma^{\mathrm{c}}_{ij}(\bR,\ii\omega)
\end{equation}
where $\Sigma^{\mathrm{c}}_{ij}(\bR,\ii\omega)$ is computed from Eq.~\eqref{eq:sigmac-ij-Rt}.

For each individual Kohn-Sham state,
\JCTC{%
Eq.~\eqref{eq:quasi-particle-equation-decomp} is solved by a damped fixed-point iteration initialized from the mean-field eigenvalue $\epsilon_{n\bk}$.
The iteration is stopped when the absolute residual of the quasi-particle equation is smaller than $10^{-5}$ Ha, and the converged value is taken as the quasi-particle energy $\epsilon^{\QP}_{n\bk}$.
This procedure usually finds the root closest to the starting mean-field state and does not perform a global search over all possible roots.
The assignment of a unique quasi-particle solution may therefore become ambiguous in multi-solution scenarios, for example in calculations of core levels.\cite{GolzeD19GW}
For the weakly correlated systems and states close to the Fermi level considered in this work, however, the self-energy is generally smooth around the quasi-particle solution, and the procedure gives stable results in most cases.
}%
Because our implementation computes the self-energy in the real-space representation, Eq.~\eqref{eq:quasi-particle-equation-decomp} can be solved for any $\bk$ point for which the Kohn-Sham reference data are available.
This places uniformly sampled $\bk$ grids and band-path $\bk$ points on the same footing.
For band-structure calculations, this procedure is equivalent to a Fourier interpolation scheme.

\subsection{Workflow}
\begin{figure}
\includegraphics[width=0.75\linewidth]{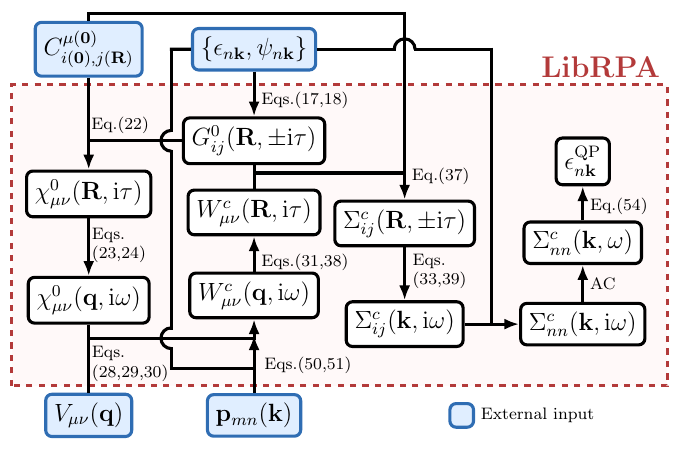}
\caption{
  Flowchart of the low-scaling $G^0W^0$ implementation in LibRPA.
  The region enclosed by the red dashed lines indicates the internal workflow in LibRPA for computing the quasi-particle energies $\epsilon^{\QP}_{n\bk}$.
  The quantities shown in blue blocks are input data provided by an external first-principles code, which is FHI-aims\cite{FHIaims25arxiv} in the present work.
  ``AC'' denotes the analytic continuation discussed in Sec.~\ref{sssec:ac}.
}
\label{fig:flowchart}
\end{figure}

The overall data flow and computational procedure of the present low-scaling $G^0W^0$ implementation are illustrated in Fig.~\ref{fig:flowchart}.
To carry out a $G^0W^0$ calculation, the required input data, including the Kohn-Sham mean-field results, LRI coefficients, Coulomb matrices, and momentum matrices, are first generated by an external first-principles code based on an NAO framework.
The momentum matrices are used to accelerate $\bk$-point convergence through the finite-size correction to the dielectric matrix described in Sec.~\ref{ssec:finite-size}.
The resulting input data are then parsed to LibRPA either through the application programming interface or through the file-based driver.
Subsequently, the intermediate quantities are evaluated according to the equations and algorithms described in the previous sections, ultimately yielding the self-energy matrix $\Sigma_{ij}(\bR)$ and the quasi-particle energies $\epsilon^{\QP}_{n\bk}$.

\subsection{Computational details}
All input data used for benchmarking the low-scaling algorithm implemented in LibRPA are generated with FHI-aims,\cite{BlumV09,FHIaims25arxiv} an all-electron, full-potential first-principles code based on numerical atomic orbitals (NAOs).
Unless stated otherwise, the real-space integration grids and basis functions are taken from the \texttt{intermediate\_gw} default species settings.
For Brillouin-zone sampling, a uniform \kgh{8} $\bk$ grid is employed for cubic unit cells unless noted otherwise.
The Perdew-Burke-Ernzerhof (PBE)\cite{PerdewJ96PBE} generalized-gradient approximation (GGA) is used for the self-consistent Kohn-Sham calculations.
To assess the present implementation in terms of both accuracy and efficiency, we also perform $G^0W^0$ calculations using the canonical $\bk$-space imaginary-frequency implementation in FHI-aims\cite{RenX21} and use the corresponding results and profiling data as reference.
In the canonical approach, the frequency convolution along the imaginary axis is carried out using modified Gauss-Legendre quadrature.
Unless stated otherwise, a 16-point Pad\'e approximant is used for analytic continuation in the canonical FHI-aims calculations, whereas in LibRPA the order of the Pad\'e approximant is set equal to the number of imaginary-frequency points.
This setup is found to work well for valence and low-lying conduction states.

In terms of test cases, we focus on extended systems with finite band gaps, namely solid-state semiconductors and insulators.
The selected materials are listed in Table~\ref{tab:lattice} in Appendix \ref{app:lattice} together with their space-group classifications and lattice parameters.
We note that spin-orbit coupling is not considered in the present work.

\section{Results and Discussions}\label{sec:results}

\subsection{Accuracy benchmark}\label{ssec:accuracy}

In this section, we assess the accuracy of our low-scaling implementation by benchmarking it against the canonical one in FHI-aims.
\JCTC{%
We emphasize that this comparison is designed to verify whether the new low-scaling implementation can reproduce the canonical one when the underlying numerical representation and shared approximations are matched.
To this end, LibRPA uses the NAO dataset generated by FHI-aims, so that the one-electron basis representation, the auxiliary RI basis, and the Coulomb matrix are shared between the two calculations.
The local RI scheme is used consistently in both implementations, and the treatment of the Coulomb singularity is aligned by applying the same head correction in the comparison.
Under this setup, the comparison isolates the low-scaling LibRPA evaluation strategy, including its use of minimax imaginary-time and imaginary-frequency grids, from the common NAO/LRI framework.
The accuracy of the underlying NAO/LRI framework as implemented in FHI-aims has been examined in the previous study on the canonical implementation.\cite{RenX21}
}%

\subsubsection{Convergence with minimax grids}\label{sssec:minimax-convergence}

\begin{figure}
\includegraphics[width=0.6\linewidth]{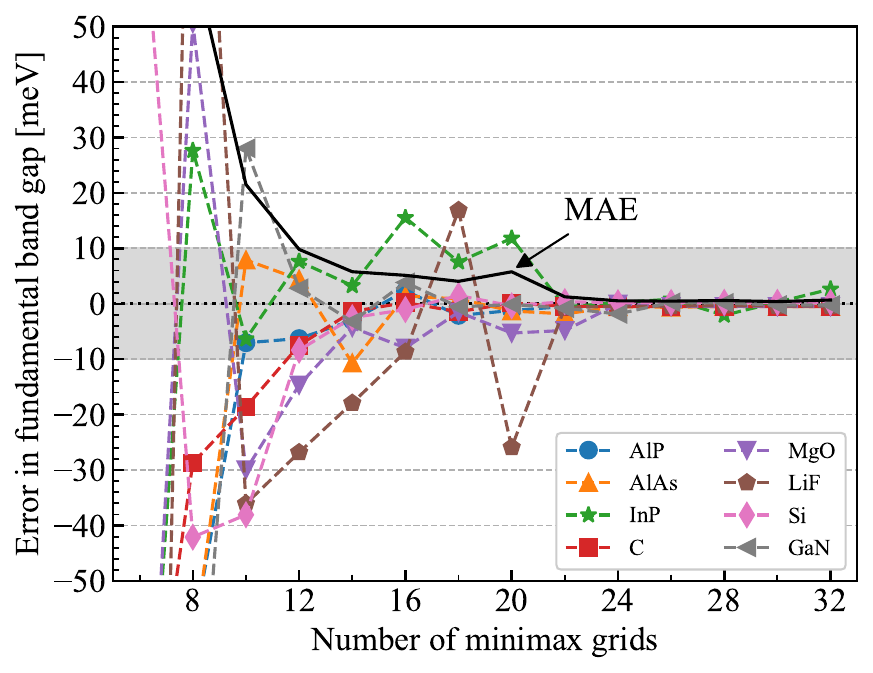}
\caption{
  Convergence of the fundamental band gaps of eight selected semiconductors and insulators with respect to the number of minimax grid points in low-scaling $G^0W^0$@PBE calculations.
  The results from the canonical implementation in FHI-aims are taken as reference.
  The shaded region indicates an error range of $\pm 10$ meV.
  The mean absolute error (MAE) over all eight materials is shown by the black solid line.
  The \texttt{intermediate\_gw} basis set and an \kgh{8} $\bk$ grid are used.
  In the canonical calculations, 60 modified Gauss--Legendre grid points are employed.
}
\label{fig:freq-conv-fund}
\end{figure}

We first examine the convergence of the quasi-particle band gap with respect to the number of minimax time/frequency grid points.
Figure \ref{fig:freq-conv-fund} shows for a set of semiconductors the error in the fundamental band gap obtained from the low-scaling implementation in LibRPA, with reference to the canonical implementation in FHI-aims based on modified Gauss-Legendre grids.
In the canonical calculations, 60 modified Gauss-Legendre grid points are used, which ensures that the reference gap values are converged to within 1 meV.
For most materials considered here, the fundamental gap from the low-scaling algorithm converges to within 10 meV of the canonical result with 16 or more minimax grid points; the main exceptions are InP and LiF, which exhibit somewhat slower convergence.
The mean absolute error (MAE) over all eight materials drops below 10 meV once 12 or more minimax points are used.
With 32 minimax grid points, the deviation in the fundamental gap is below 5 meV for most systems, which is substantially smaller than the typical uncertainty associated with the convergence of the orbital basis set and BZ sampling.\cite{JiangH16,ZhangMY19IBX}

\subsubsection{Quasi-particle band structures}

\begin{figure}
\includegraphics[width=0.85\linewidth]{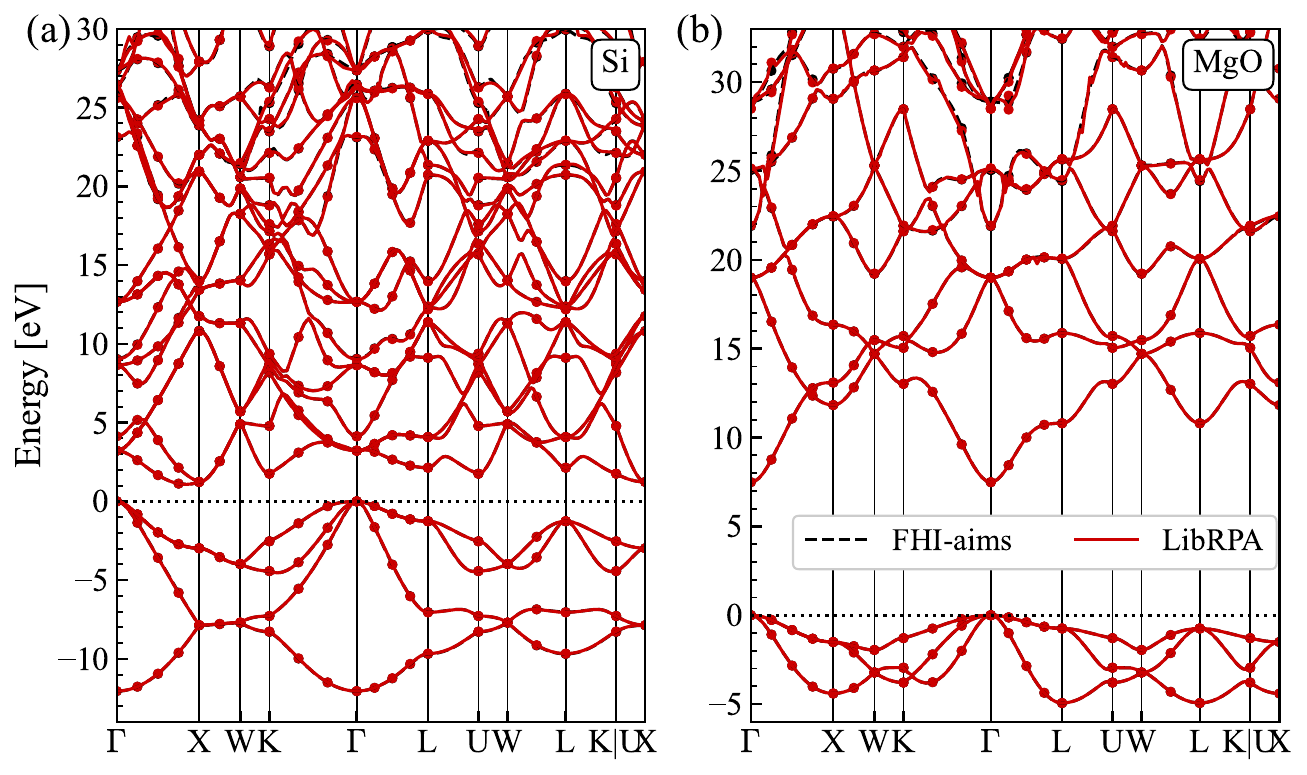}
\caption{
  $G^0W^0$@PBE quasi-particle band structures of (a) Si and (b) MgO computed
  by the canonical implementation in FHI-aims (black) and the low-scaling implementation in LibRPA (red).
  Dots indicate the quasi-particle energies evaluated on the uniform $\bk$-grid.
  The FHI-aims calculations use 60 modified Gauss-Legendre grid points, while the LibRPA calculations use 32 minimax time/frequency points.
  For both materials, the band structures are aligned by the valence band maximum as the energy zero.
  The \texttt{intermediate\_gw} basis set and an \kgh{8} $\bk$ grid are used.
}
\label{fig:bandstructure}
\end{figure}

To further validate the implementation, Figure~\ref{fig:bandstructure} compares the $G^0W^0$ quasi-particle band structures of Si and MgO obtained from the canonical implementation in FHI-aims and the low-scaling implementation in LibRPA.
For both systems, the two band structures are nearly indistinguishable in the vicinity of the Fermi level, and the agreement remains excellent up to approximately 30 eV above the valence band maximum (VBM).
We note that in the canonical implementation of FHI-aims, the quasi-particle energies along the chosen $\bk$-path are evaluated explicitly and are therefore free of interpolation errors.
The close agreement therefore demonstrates the reliability of the LibRPA implementation for $G^0W^0$ band-structure calculations, provided that a sufficiently dense uniform $\bk$ grid is used for the underlying self-energy calculation.
In the present examples (cubic unit cells), an \kgh{8} $\bk$ grid is sufficient to achieve this accuracy.

\subsubsection{Benchmark across semiconductors and insulators}

We next benchmark the low-scaling $G^0W^0$ implementation across a broader set of prototypical semiconductors and insulators.
The resulting fundamental band gaps are summarized in Table~\ref{tab:band-gaps}.
For the 24 gapped systems considered here, the difference between the canonical and low-scaling $G^0W^0$ gaps is below 5 meV for most materials when 32 minimax grid points are used.

Noticeable exceptions are GaAs (24.8 meV), GaSb (14.3 meV), and ZnO (24.1 meV), with GaAs in particular showing comparatively slow convergence with respect to the minimax grid size.
\JCTC{%
Additional analysis for GaAs shows that this slow convergence is caused by a self-energy pole located close to the quasi-particle solution in the conduction-band minimum (CBM), which makes the extracted CBM energy highly sensitive to the frequency grid and the analytic continuation.
For weakly correlated systems, such a nearby pole is likely an artifact at the $G^0W^0$@PBE level and may give rise to spurious multiple solutions and artificial spectral-weight transfer from the quasi-particle peak to satellite features.
It has been shown that eigenvalue self-consistency, for example in ev$GW$,\cite{ShishkinM07PRB} or the Hedin-shift technique,\cite{PollehnTJ98} can alleviate this artifact by shifting the pole away from the quasi-particle solution, thereby recovering a single physical solution and improving frequency-grid convergence.\cite{SchambeckM24,PasquierR25}
We also verified that, with an improved starting point such as PBE0, one-shot $GW$ for GaAs yields clear single quasi-particle solutions for both the VBM and CBM, and that the corresponding quasi-particle energies are converged within 1 meV already at 16 minimax grid points.
}%

\JCTC{%
Aside from these challenging cases, the agreement between the canonical and low-scaling results remains consistent with the convergence trend discussed in Sec.~\ref{sssec:minimax-convergence}, confirming that the low-scaling implementation reproduces the canonical results for a wide range of materials.
}%
The comparison of $G^0W^0$ band gaps computed in the FHI-aims NAO basis against experiment and against results from other codes and implementations has been discussed in Ref.~\citenum{RenX21} and more recently in the cross-code benchmark study of Ref.~\citenum{AziziM25}.

\begin{table}[!ht]
\centering
\caption{
  Fundamental band gaps $E^{\mathrm{fund}}_{\mathrm{g}}$ of 24 semiconductors computed at the PBE and $G^0W^0$@PBE levels.
  For $G^0W^0$, results from the canonical implementation in FHI-aims and the low-scaling implementation in LibRPA are compared.
  For each material, the difference between the two $G^0W^0$ gaps is defined as
  $\Delta = \abs{E^{\mathrm{fund}}_{\mathrm{g}}(\mathrm{FHI\text{-}aims}) - E^{\mathrm{fund}}_{\mathrm{g}}(\mathrm{LibRPA})}$.
  In FHI-aims, 60 modified Gauss-Legendre grid points are used, while in LibRPA 32 minimax grid points are employed.
  A uniform \kgh{8} $\bk$ grid is used throughout.
  An asterisk (*) attached to the material name indicates that the corresponding FHI-aims result is obtained using a 32-point Pad\'e approximant.
  Experimental values (``Expt.'') are taken from Ref.~\citenum{RenX21} unless specified otherwise. All values are given in eV.
}
\label{tab:band-gaps}
\begin{threeparttable}
\begin{tabular}{lrrccc}
\toprule
System & Expt. & PBE & $G^0W^0$ (FHI-aims) & $G^0W^0$ (LibRPA) & $\Delta$ \\
\midrule
AlP  &  2.53 & 1.5706 &  2.2859 &  2.2854 & 0.0005 \\
AlAs &  2.27 & 1.4399 &  2.0438 &  2.0433 & 0.0005\\
BN   &  6.66 & 4.5017 &  6.4277 &  6.4272 & 0.0005\\
BP   &  2.2  & 1.2470 &  2.0092 &  2.0090 & 0.0002\\
C    &  5.85 & 4.1833 &  5.8206 &  5.8201 & 0.0005\\
CaO  &  6.4\tnote{a} & 3.6283 &  6.2930 &  6.2889 & 0.0041\\
CdS  &  --   & 1.1704 &  2.0447 &  2.0456 & 0.0009\\
GaN  &  3.64 & 1.6862 &  2.9718 &  2.9715 & 0.0003\\
GaP  &  2.43 & 1.6022 &  2.2234 &  2.2237 & 0.0003\\
GaAs*&  1.42\tnote{b} & 0.5427 &  1.2997 &  1.2749 & 0.0248\\
GaSb &  0.81\tnote{b} & 0.1288 &  0.6831 &  0.6974 & 0.0143\\
InP  &  1.34\tnote{b} & 0.7149 &  1.2727 &  1.2753 & 0.0026\\
MgO  &  7.98 & 4.7250 &  7.4892 &  7.4887 & 0.0005\\
MgS  &  4.5\tnote{b}  & 2.7851 &  4.2469 &  4.2464 & 0.0005\\
Si   &  1.23 & 0.5758 &  1.0802 &  1.0805 & 0.0003\\
SiC  &  2.57 & 1.3799 &  2.4589 &  2.4584 & 0.0005\\
LiF  & 14.48 & 9.1746 & 13.8336 & 13.8332 & 0.0004\\
LiCl &  9.8  & 6.3124 &  8.8083 &  8.8078 & 0.0005\\
NaF  &  11.7\tnote{c} & 6.3080 & 10.8456 & 10.8452 & 0.0004\\
NaCl &  8.6  & 5.0889 &  7.9100 &  7.9095 & 0.0005\\
KF   &  10.96\tnote{d}& 6.1235 &  9.8894 &  9.8890 & 0.0004\\
KCl  &  8.65\tnote{e} & 5.1649 &  7.9367 &  7.9362 & 0.0005\\
ZnS  &  3.68\tnote{b} & 2.0915 &  3.2893 &  3.2889 & 0.0004\\
ZnO  &   --  & 0.6797 &  2.0462 &  2.0221 & 0.0241\\
\bottomrule
\end{tabular}
\begin{tablenotes}[flushleft]
\footnotesize
\item[a] Ref.~\citenum{LushchikC94}.
\item[b] Ref.~\citenum{JiangH16}.
\item[c] Ref.~\citenum{NakaiS69}.
\item[d] Ref.~\citenum{TomikiT69}.
\item[e] Ref.~\citenum{CasalboniM91}.
\end{tablenotes}
\end{threeparttable}
\end{table}

\subsection{Efficiency benchmark}\label{ssec:efficiency}

In this section, we demonstrate the efficiency of our low-scaling implementation, focusing on the role of matrix filtering and scaling with system size.

\subsubsection{Filtering of real-space matrices}

Due to the spatial locality of NAOs and the represented physical quantities in real space, the sparsity of the relevant matrices can be exploited to speed up the computation.
In this work, we exploit the sparsity at the level of atom-pair matrix or tensor blocks.
For each type of matrix or tensor, a filtering threshold is defined and applied at the time when the corresponding blocks are distributed, so that blocks do not need to be checked repeatedly during the subsequent computation.
A block is discarded entirely if the maximum absolute value of its elements is smaller than the threshold.
This block-wise treatment differs from the conventional element-wise handling in sparse-matrix algorithms.

To assess how atom-pair block filtering affects the computational efficiency of the present $G^0W^0$ implementation, we apply it to the main quantities entering the calculation, namely the response function $\chi^0$, the exact-exchange operator $\Sigma^{\mathrm{x}}$, and the correlation self-energy $\Sigma^{\mathrm{c}}$.
For convenience, we denote by $\eta_A(T)$ the threshold applied to input blocks of quantity $A$ when computing a target quantity $T$.
The relevant thresholds are therefore $\eta_C(\chi)$ and $\eta_G(\chi)$ for the RI coefficients $C$ and the Green's function $G$ in the evaluation of $\chi^0_{\mu\nu}$ [Eq.~\eqref{eq:chi0-mn}];
$\eta_C(\Sigma^{\mathrm{x}})$, $\eta_V(\Sigma^{\mathrm{x}})$, and $\eta_D(\Sigma^{\mathrm{x}})$ for $C$, the bare Coulomb matrix $V$, and the density matrix $D$ in the evaluation of $\Sigma^{\mathrm{x}}$ [Eq.~\eqref{eq:vexx-R}];
and $\eta_C(\Sigma^{\mathrm{c}})$, $\eta_W(\Sigma^{\mathrm{c}})$, and $\eta_G(\Sigma^{\mathrm{c}})$ for $C$, the screened Coulomb interaction $\Wc$, and $G$ in the evaluation of $\Sigma^{\mathrm{c}}$ [Eq.~\eqref{eq:sigmac-ij-Rt}].
In the present work, the thresholds for $\Sigma^{\mathrm{x}}$ are fixed at conservative default values when other thresholds are varied, because the computational cost of $\Sigma^{\mathrm{x}}$ is one to two orders of magnitude smaller than that of $\Sigma^{\mathrm{c}}$ (see Sec.~\ref{sssec:contraction-sigc}), and so its impact on the overall performance is negligible.

\begin{figure}
\includegraphics[width=0.95\linewidth]{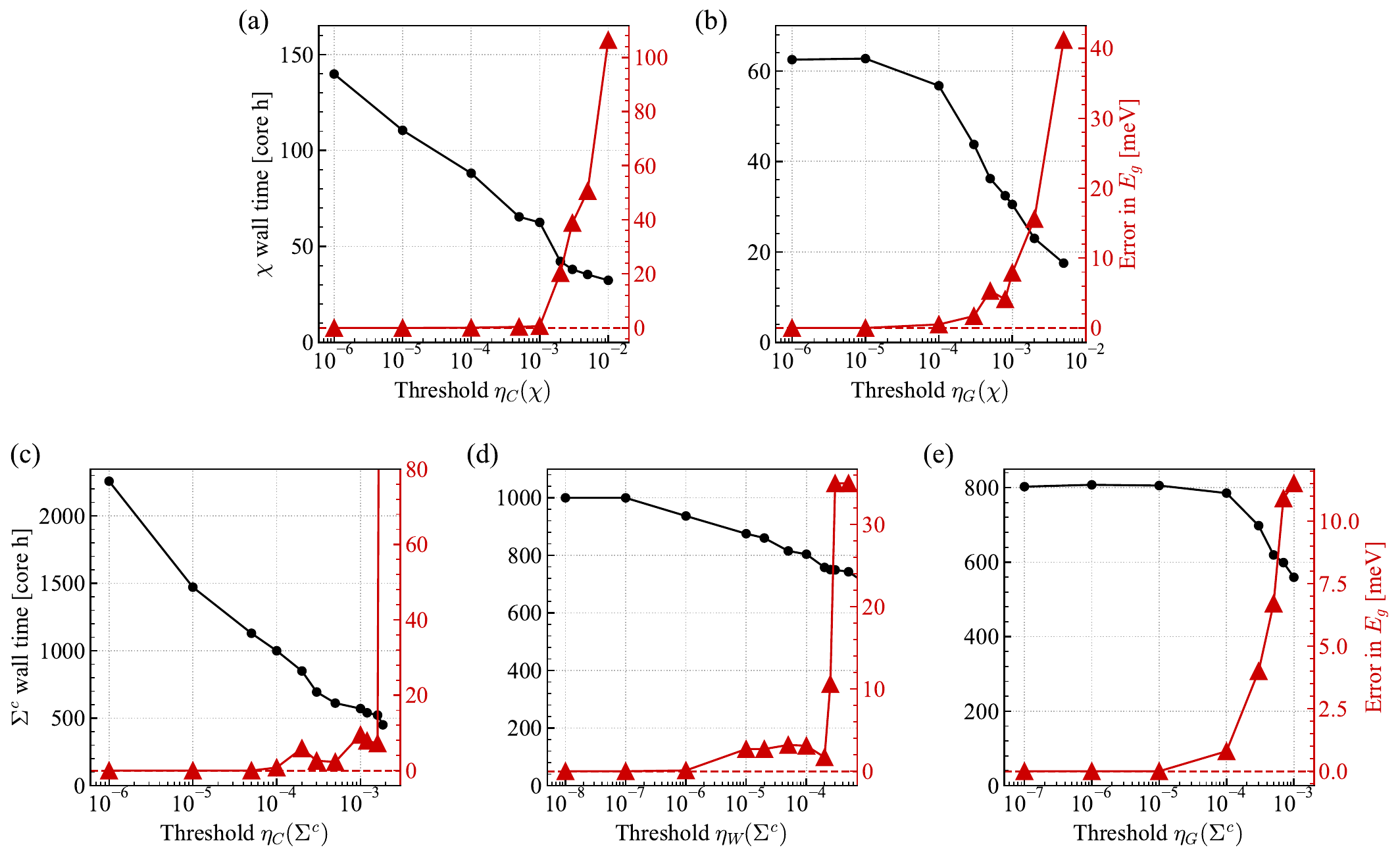}
\caption{
  Error in the $G^0W^0$@PBE fundamental gap $E_{\mathrm{g}}$ of diamond, computed with the low-scaling algorithm using different filtering thresholds (red triangles, in meV):
  (a) RI coefficients $\eta_C$ and (b) Green's function $\eta_G$ for the response function $\chi^0$;
  (c) RI coefficients $\eta_C$, (d) screened Coulomb matrix $\eta_W$, and (e) Green's function $\eta_G$ for the correlation self-energy $\Sigma^{\mathrm{c}}$.
  The result obtained with the smallest tested threshold is taken as the reference.
  See the main text for the details of the threshold setup.
  Black circles indicate the wall time (in core hours) for $\chi^0_{\mu\nu}$ [Eq.~\eqref{eq:chi0-mn}] in the upper panels and for $\Sigma^{\mathrm{c}}_{ij}$ [Eq.~\eqref{eq:sigmac-ij-Rt}] over all imaginary-time points in the lower panels.
  The calculations use 16 minimax grid points and MPI+OpenMP parallelization (4 MPI tasks, each with 48 threads) on two Intel Xeon(R) Platinum 8468 nodes with 96 cores each.
}
\label{fig:matrix_screening}
\end{figure}

The impact of matrix filtering on the performance of $G^0W^0$ for the diamond unit cell of carbon with an \kgh{8} $\bk$ grid (corresponding to an  \kgh{8} BvK supercell in real space) is shown in Fig.~\ref{fig:matrix_screening}.
The error in the quasi-particle band gap is measured relative to the reference result obtained with the smallest tested threshold.

We first examine the thresholds for the response function $\chi^0$, shown in the upper panel of Fig.~\ref{fig:matrix_screening}.
As shown in Fig.~\ref{fig:matrix_screening}(a), $\eta_C(\chi)$ can be increased to $10^{-3}$ while introducing only a 0.7 meV error in the band gap.
At the same time, the computation of $\chi^0(\ii\tau)$ is accelerated by a factor of 2.24 relative to the reference with $\eta_C(\chi)=10^{-6}$.
Fixing $\eta_C(\chi)=10^{-3}$, we then increase $\eta_G(\chi)$ [Fig.~\ref{fig:matrix_screening}(b)].
The computational time starts to decrease for $\eta_G(\chi)\gtrsim 10^{-5}$.
At $\eta_G(\chi)=10^{-4}$, the band-gap error is 0.5 meV and the speedup is 1.10.
Further increasing $\eta_G(\chi)$ leads to band-gap errors exceeding 1 meV.

Independent tests of matrix filtering for the correlation self-energy are shown in the lower panel of Fig.~\ref{fig:matrix_screening}.
Increasing $\eta_C(\Sigma^{\mathrm{c}})$ [Fig.~\ref{fig:matrix_screening}(c)] yields a substantial speedup.
At $\eta_C(\Sigma^{\mathrm{c}})=10^{-4}$, the calculation is accelerated by a factor of 2.26, while the band-gap error remains 0.8 meV relative to the reference with $\eta_C(\Sigma^{\mathrm{c}})=10^{-6}$.
Upon further increasing $\eta_C(\Sigma^{\mathrm{c}})$, the error becomes non-monotonic.
In particular, using $\eta_C(\Sigma^{\mathrm{c}})=10^{-3}$, which is optimal for $\eta_C(\chi)$, gives a speedup of 3.95 but increases the band-gap error to 9.5 meV.
Although the error remains below 10 meV for $\eta_C(\Sigma^{\mathrm{c}}) < 1.6\times10^{-3}$, it rises sharply to more than 400 meV beyond this point.
We therefore adopt the conservative choice $\eta_C(\Sigma^{\mathrm{c}})=10^{-4}$ in the following tests.

The stronger speedup of $\Sigma^{\mathrm{c}}$ under $\mathbf{C}$ filtering can be traced mainly to the term $\boldsymbol{\Sigma}^{\mathrm{c},\mathrm{(D)}}$, which is also the dominant contribution in practice.
As shown in Fig.~\ref{fig:sigmac-rearrange-pattern}(d), this contribution is accumulated into $\boldsymbol{\Sigma}^{\mathrm{c}}_{KL}$, where the target atom pair $(K,L)$ is determined by the two RI-coefficient tensors.
Consequently, if either side is removed by $\mathbf{C}$ filtering, the corresponding contribution vanishes entirely.
Denoting by $p$ the retained fraction of $\mathbf{C}$ blocks after filtering, the computational effort of this term therefore decreases approximately with the joint survival probability of two $\mathbf{C}$-dependent branches, leading to a speedup close to the observed $1/p^2$ trend.
This mechanism does not apply equally to the other terms, nor to $\chi^0$, which explains their weaker sensitivity to $\mathbf{C}$ filtering.

With $\eta_C(\Sigma^{\mathrm{c}})=10^{-4}$ fixed, we further examine $\eta_W(\Sigma^{\mathrm{c}})$ [Fig.~\ref{fig:matrix_screening}(d)] and $\eta_G(\Sigma^{\mathrm{c}})$ [Fig.~\ref{fig:matrix_screening}(e)], the latter additionally with $\eta_W(\Sigma^{\mathrm{c}})=10^{-4}$.
In both cases, no speedup larger than 1.2 is obtained while keeping the band-gap error below 1 meV.
Accordingly, for the following scaling tests we adopt a conservative set of filtering thresholds, namely $\eta_C(\chi)=10^{-3}$ and $\eta_C(\Sigma^{\mathrm{x}})=\eta_C(\Sigma^{\mathrm{c}})=10^{-4}$.

We emphasize that the optimal threshold combination and speedup factor in general depend on the system type and size. The present choice is intended to maximize transferability while maintaining a balance between efficiency and accuracy. For larger system sizes (bigger supercells or denser $\bk$ grids), we expect a higher speedup factor with the present choice of filtering thresholds.

\subsubsection{Scaling with number of k-points}
\begin{figure}
\includegraphics[width=0.65\linewidth]{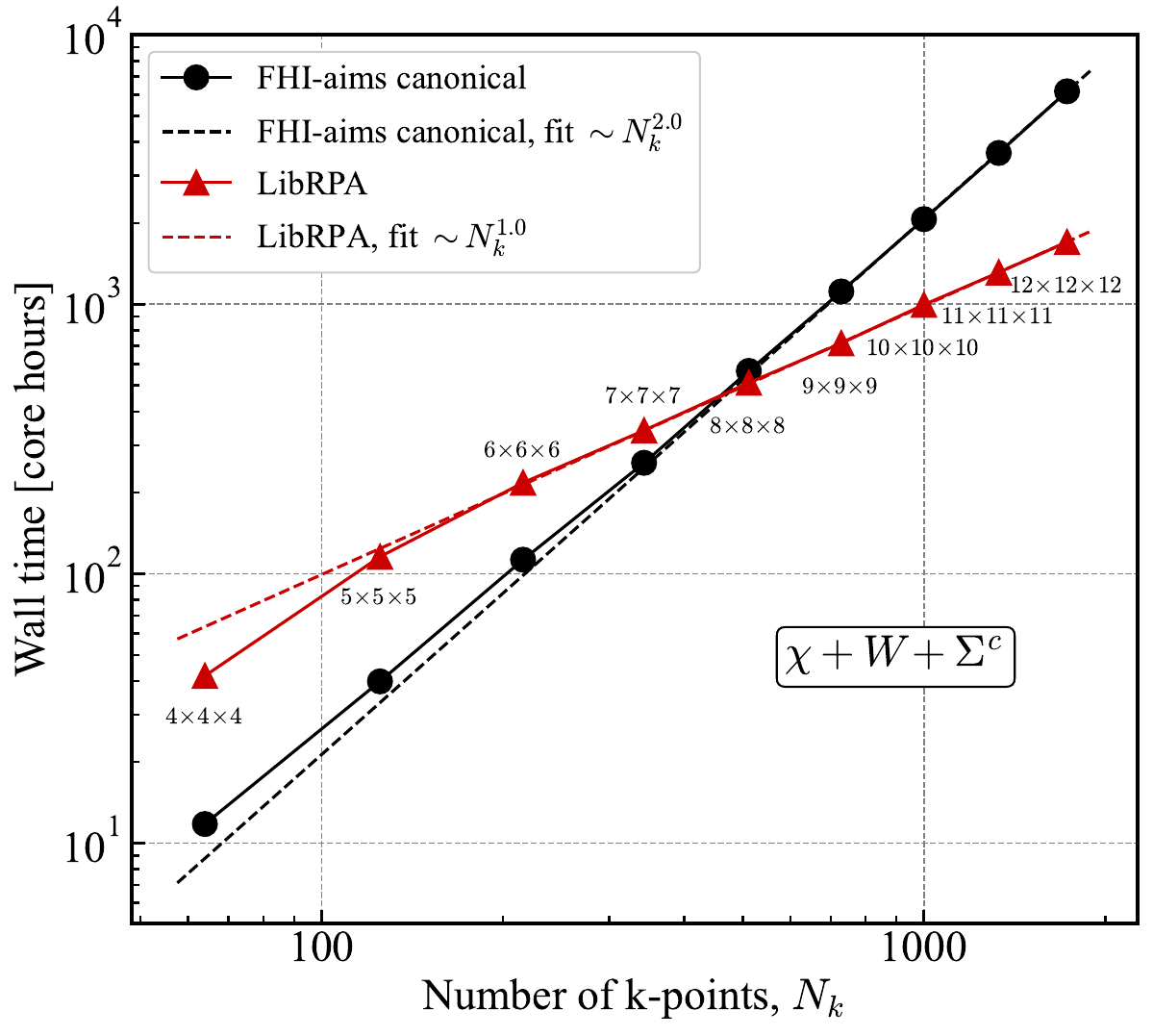}
\caption{\label{fig:k-scaling}
  Wall-time scaling (in core hours) of $G^0W^0$@PBE calculations for silicon with respect to the number of $\bk$ points in the first Brillouin zone,
  comparing the canonical implementation in FHI-aims (black) and the low-scaling implementation in LibRPA (red).
  The reported wall time includes the computation of $\chi^0$, the construction of $\Wc$ from $\chi^0$, and the evaluation of $\Sigma^{\mathrm{c}}$.
  The \texttt{intermediate\_gw} species default from FHI-aims is used.
  \JCTC{Both calculations employ 16-point imaginary-frequency grids.}
  The LibRPA calculations use filtering thresholds of $\eta_C(\chi)=10^{-3}$ and $\eta_C(\Sigma^{\mathrm{x}})=\eta_C(\Sigma^{\mathrm{c}})=10^{-4}$.
  The calculations are performed with MPI+OpenMP parallelization (16 MPI tasks, each with 24 threads) on four Intel Xeon(R) Platinum 8468 nodes with 96 cores each. 
}
\end{figure}

The scaling of the computational time with respect to the number of $\bk$ points is benchmarked for the silicon unit cell.
The filtering thresholds are fixed to the conservative values determined above, namely $\eta_C(\chi)=10^{-3}$ and $\eta_C(\Sigma^{\mathrm{x}})=\eta_C(\Sigma^{\mathrm{c}})=10^{-4}$.
Figure~\ref{fig:k-scaling} reports the total wall times for computing $\chi^0(\ii\tau)$, constructing $\Wc$ from $\chi^0$, and evaluating $\Sigma^{\mathrm{c}}(\ii\tau)$.
The canonical implementation reproduces the expected $\mathcal{O}(\Nk^2)$ scaling, whereas the present low-scaling algorithm shows clear linear scaling for $\Nk>216$ (\kgh{6}).
Although the low-scaling implementation has a larger prefactor and is therefore less favorable at small $\bk$ meshes, it becomes increasingly advantageous beyond the crossover region at intermediate $\bk$ sampling, which in the present case lies between \kgh{7} and \kgh{8}.

\subsubsection{Scaling with system size}

In this section, we analyze the computational scaling of the present low-scaling algorithm in LibRPA with respect to system size and compare it with that of the canonical implementation in FHI-aims.

\begin{figure}
\includegraphics[width=0.9\linewidth]{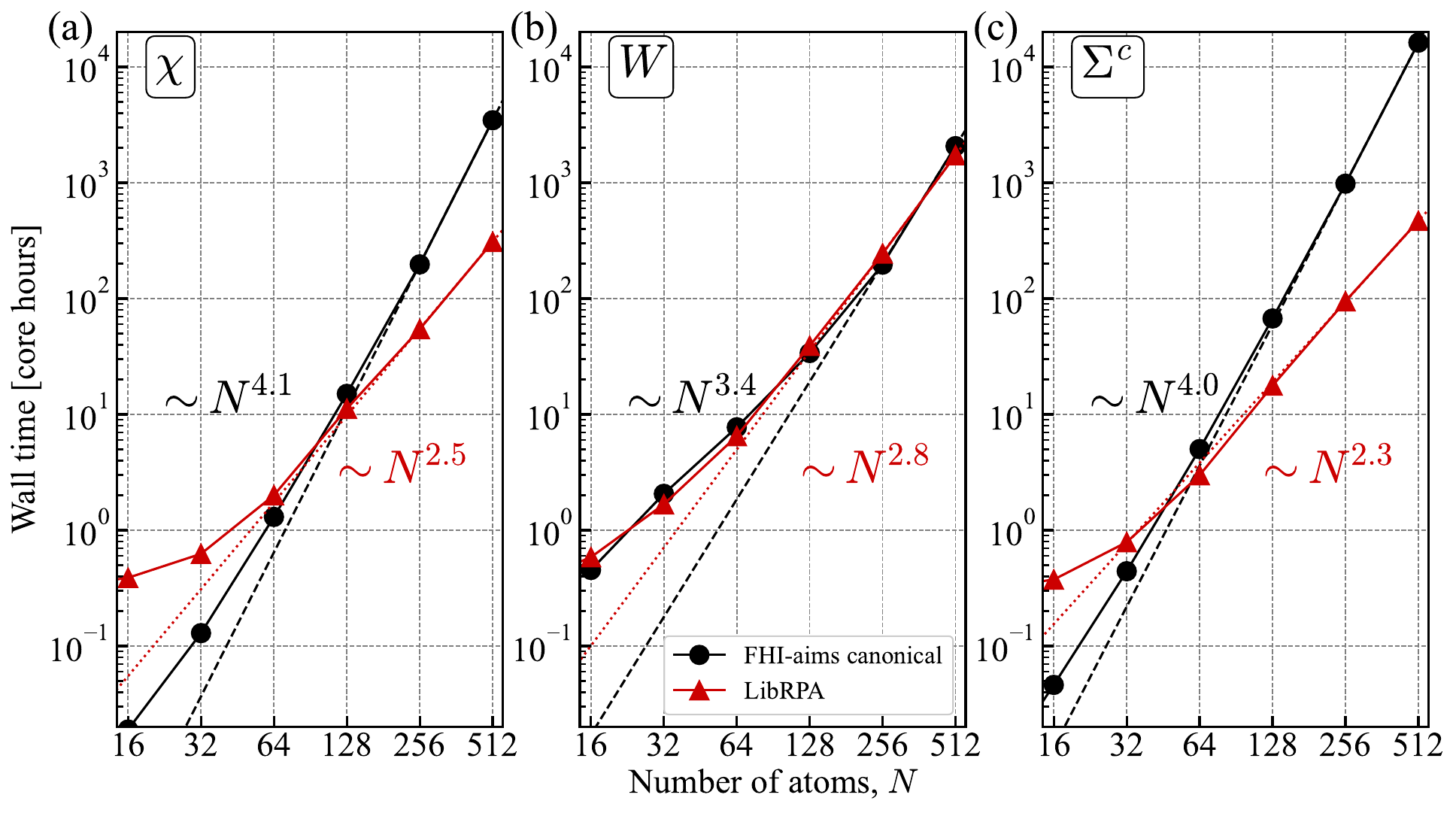}
\caption{
  Wall-time scaling (in core hours) of $G^0W^0$@PBE calculations for diamond with respect to the number of atoms in the unit cell,
  comparing the canonical implementation in FHI-aims (black circles) and the low-scaling implementation in LibRPA (red triangles).
  The reported wall times correspond to (a) $\chi^0$, (b) the construction of $\Wc$ from $\chi^0$, and (c) $\Sigma^{\mathrm{c}}$.
  The \texttt{light} species default\JCTC{, 16-point imaginary-frequency grids,} and a single $\Gamma$ point are used.
  The calculations are performed with the same computational setup as in Fig.~\ref{fig:k-scaling}.
}
\label{fig:scaling-individual}
\end{figure}

We first examine the three most time-consuming components for large atomistic systems, namely the evaluation of $\chi^0$, the construction of $\Wc$ [Eqs.~\eqref{eq:symeps} and \eqref{eq:w0-mu-qw-symeps}], and the evaluation of $\Sigma^{\mathrm{c}}$.
The corresponding profiling results for diamond supercells of carbon containing up to 512 atoms are shown in Fig.~\ref{fig:scaling-individual}.
We note that, although a $\Gamma$-only supercell and a primitive-cell calculation with the corresponding $\bk$ grid represent the same BvK supercell, their computational costs are not equivalent.
In the $\Gamma$-only supercell calculation, translational symmetry within the BvK cell is no longer exploited explicitly, and the loops must traverse all atom pairs in the supercell.
In the primitive-cell calculation, by contrast, atom pairs related by lattice translations are treated as equivalent and represented by the same real-space block.
Consequently, for a BvK cell containing $\Nk$ primitive cells, the supercell formulation must store and process roughly $\Nk$ times more distinct blocks than the corresponding $\bk$-grid formulation. The $\Gamma$-only supercell calculations will be needed for complex materials when the translational symmetry is lost.

We begin with the response function [Fig.~\ref{fig:scaling-individual}(a)].
The canonical implementation in FHI-aims exhibits the expected quartic scaling, i.e., approximately $\mathcal{O}(N^{4})$.
By contrast, the present low-scaling algorithm in LibRPA shows an empirical scaling of $\mathcal{O}(N^{2.5})$, which is slightly steeper than the asymptotic quadratic behavior expected from the formal analysis.
This indicates that, for the system sizes considered here, the neighbor list cannot yet be regarded as effectively invariant with increasing supercell size.
Nevertheless, the reduced scaling already leads to a crossover at around 80 atoms for $\chi^0$.

For constructing $\Wc$ from $\chi^0$ [Fig.~\ref{fig:scaling-individual}(b)], the wall times in FHI-aims and LibRPA are very similar.
This is because this step is dominated in both implementations by dense-matrix operations, in particular matrix multiplication and inversion, which scale cubically with system size. The observed scaling is consistent with this expectation.

For $\Sigma^{\mathrm{c}}$ [Fig.~\ref{fig:scaling-individual}(c)],
the canonical implementation in FHI-aims again shows the expected quartic scaling,
whereas the low-scaling implementation in LibRPA exhibits an empirical scaling of $\mathcal{O}(N^{2.3})$.
The crossover occurs at around 40 atoms.
For supercells with $N \gtrsim 256$, the computational cost of $\Sigma^{\mathrm{c}}$ in the canonical implementation is already about one order of magnitude larger.
We note that in FHI-aims, only the diagonal matrix elements in the Kohn-Sham basis, $\Sigma^{\mathrm{c}}_{nn}(\bk)$, are computed by default for states near the Fermi level.
Computing the full self-energy matrix would therefore be even more expensive.
By contrast, the low-scaling implementation in LibRPA computes the self-energy in the NAO basis and subsequently transforms it unitarily to the Kohn-Sham basis, so that the full matrix $\Sigma^{\mathrm{c}}_{mn}(\bk)$ is readily available.

\begin{figure}
\includegraphics[width=0.55\linewidth]{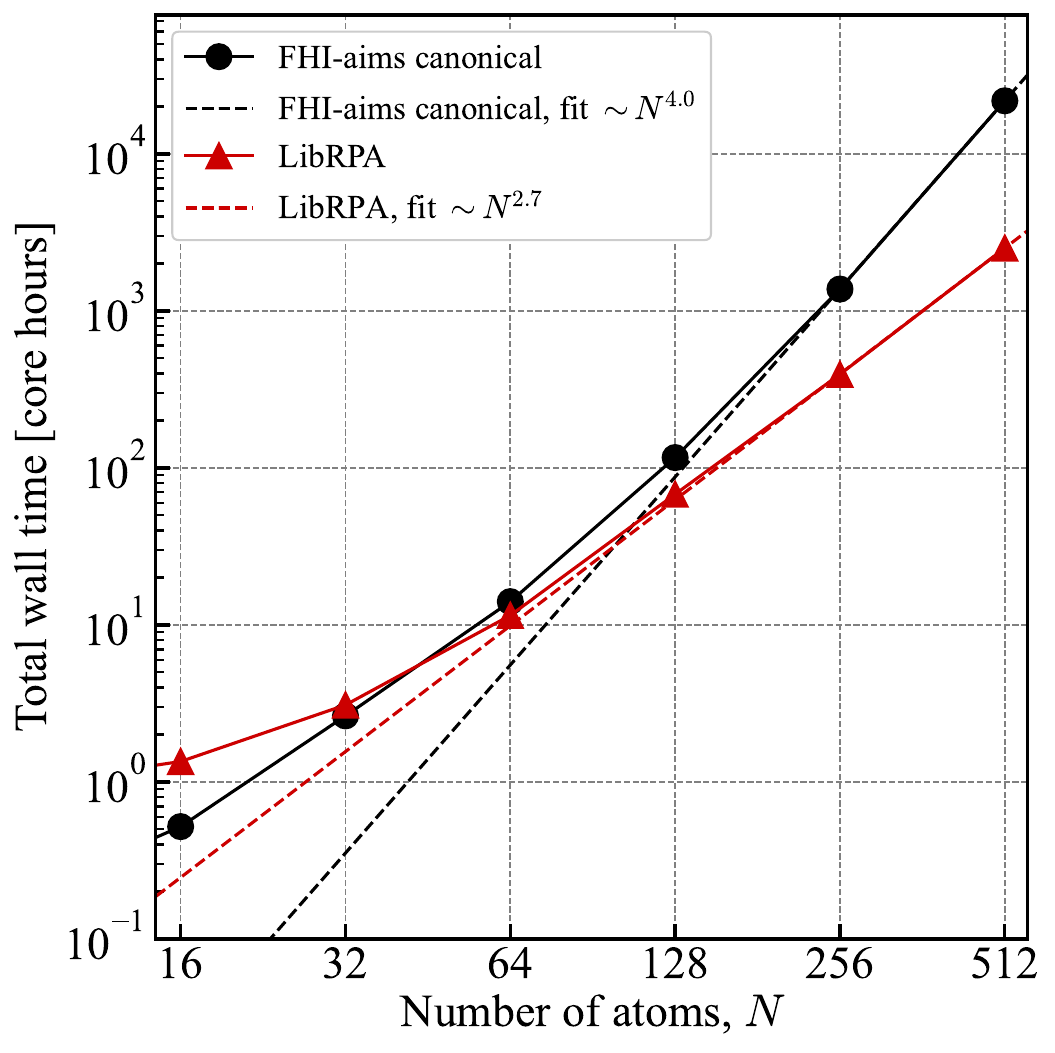}
\caption{
  Same as Fig.~\ref{fig:scaling-individual}, but for the total computer time summed over the three components.
}
\label{fig:scaling-total}
\end{figure}

Summing up the three contributions, the overall wall-time scaling with system size is shown in Fig.~\ref{fig:scaling-total}.
For the canonical reciprocal-space algorithm implemented in FHI-aims, the total cost scales quartically with the number of atoms in the unit cell, consistent with the dominant contribution from $\Sigma^{\mathrm{c}}$ shown in Fig.~\ref{fig:scaling-individual}.
For the present low-scaling algorithm in LibRPA, an empirical scaling of $\mathcal{O}(N^{2.7})$ is observed.
This intermediate exponent reflects the fact that the construction of $\Wc$ already dominates the total cost in the tested size range, while $\chi^0$ and $\Sigma^{\mathrm{c}}$ still contribute appreciably.
As the system size increases further, both $\chi^0$ and $\Sigma^{\mathrm{c}}$ are expected to approach quadratic scaling,
so the construction of $\Wc$, which remains cubic, will become the main bottleneck in the low-scaling $G^0W^0$ workflow.

\subsubsection{Strong scaling}

Efficient utilization of modern massively parallel architectures is essential for treating large-scale systems.
This requires that the calculation remains efficient on thousands of CPU cores.
LibRPA employs a hybrid MPI+OpenMP parallelization scheme and can therefore exploit multiple compute nodes effectively.

\begin{figure}
\includegraphics[width=0.55\linewidth]{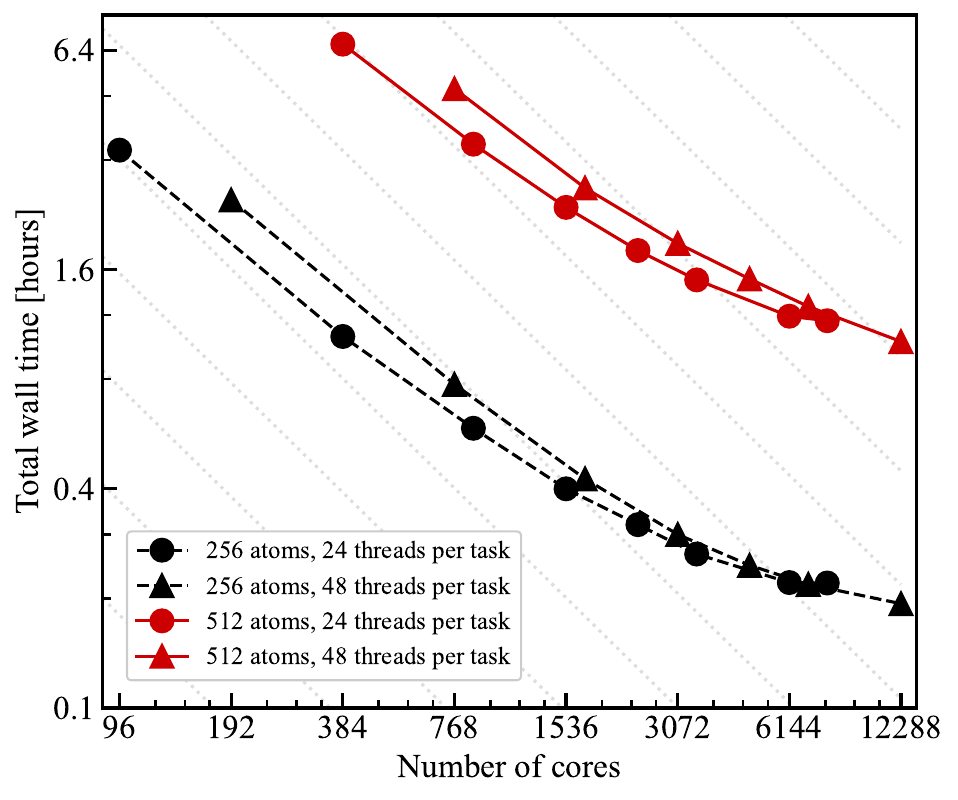}
\caption{
  Strong scaling of the $G^0W^0$@PBE calculation in LibRPA for diamond supercells of carbon.
  Results are shown for the 256-atom (black dashed) and 512-atom (red solid) supercells.
  Calculations using 24 and 48 threads per MPI task are represented by circles and triangles, respectively.
  \JCTC{The computational setup and node specifications are the same as in Fig.~\ref{fig:scaling-individual}.}
}
\label{fig:strong-scaling}
\end{figure}

Figure~\ref{fig:strong-scaling} shows the strong scaling of LibRPA for diamond supercells of carbon containing 256 and 512 atoms.
For the 256-atom supercell, the calculation with 24 threads per MPI task scales up to 81 nodes (7776 cores), yielding a speedup of 16.7 relative to the single-node run and a parallel efficiency of 20.6\%.
With 48 threads per MPI task, the same system scales up to 128 nodes (12288 cores), with a speedup of 12.8 relative to the two-node run and a parallel efficiency of 20.0\%.
At a fixed total number of CPU cores, the calculation generally favors a larger number of MPI tasks over heavier threading.
This is mainly because the construction of $\Wc$ is based on ScaLAPACK and benefits more strongly from MPI parallelization.
The 512-atom supercell exhibits a similar scaling limit and comparable parallel efficiency for both threading configurations, indicating that the present implementation remains scalable as the system size increases.

Overall, these results demonstrate that the present low-scaling $G^0W^0$ implementation can be efficiently applied on the order of $10^4$ CPU cores for systems of this size.
The current parallel efficiency is limited mainly by the evaluation of the Green's function, which has not yet been optimized to the same extent as the tensor-contraction kernels discussed above.

\section{Concluding Remarks}\label{sec:conclude}

In this work, we have presented an all-electron low-scaling $G^0W^0$ implementation for periodic systems based on localized atomic orbitals, using the LibRPA framework interfaced with FHI-aims.
Based on the space-time formalism and the LRI approximation, the method reformulates the key many-body quantities in terms of real-space tensor contractions over atom-pair blocks, thereby enabling efficient exploitation of the sparsity induced by basis locality.
We have developed optimized contraction algorithms for the irreducible response function $\chi^0$, the exchange self-energy $\Sigma^{\mathrm{x}}$, and the correlation self-energy $\Sigma^{\mathrm{c}}$, together with the corresponding workflow for periodic $G^0W^0$ calculations.

Benchmark calculations demonstrate that the present implementation reproduces the canonical periodic $G^0W^0$ results in FHI-aims with high fidelity.
For a representative set of semiconductors and insulators, quasi-particle band gaps and band structures obtained from the low-scaling implementation agree closely with those from the canonical approach, with meV-level differences for most systems when sufficiently converged minimax grids are employed.
We further show that real-space matrix filtering based on atom-pair blocks can provide substantial additional speedup while maintaining controlled accuracy.
In particular, filtering of the RI-coefficient blocks is especially effective for the correlation self-energy.

Using a conservative set of filtering thresholds, the efficiency benchmarks confirm that the present implementation exhibits the expected reduction in scaling.
With respect to $\bk$-point sampling, whereas the canonical implementation follows the expected quadratic scaling, the low-scaling implementation reaches linear scaling at sufficiently dense $\bk$ meshes (\kgh{6} in the silicon test) and becomes advantageous beyond the crossover region, which lies between \kgh{7} and \kgh{8} in this case.
With respect to system size, the low-scaling implementation reduces the cost of the dominant response-function and self-energy steps to near-quadratic behavior.
For diamond supercells containing up to 512 atoms, empirical scalings of $\mathcal{O}(N^{2.5})$ and $\mathcal{O}(N^{2.3})$ are observed for $\chi^0$ and $\Sigma^{\mathrm{c}}$, respectively.
As a result, the overall scaling is reduced to approximately $\mathcal{O}(N^{2.7})$ for the systems considered here.
In addition, the implementation shows good strong scaling on modern massively parallel architectures up to the order of $10^4$ CPU cores.
Although demonstrated here in an NAO-based implementation, the underlying low-scaling framework is general and can in principle be extended to other localized AO schemes.

Future work can be categorized into further optimization of the present implementation and extensions to new physical regimes.
On the optimization side, improving memory handling will be important for pushing the accessible system size further and for more fully reaching the asymptotic quadratic-scaling regime of the low-scaling response-function and self-energy algorithms.
Another key target is the construction of $\Wc$, which becomes the main bottleneck for large systems.
Because this step is dominated by dense linear-algebra operations, GPU acceleration is expected to be an effective route for further improvement.

On the extension side, several directions are particularly attractive.
Although the code base already supports $GW$ calculations including SOC, a systematic study within the present low-scaling periodic framework remains to be carried out.
A more complete treatment of the $\Gamma$ point is also needed, including wing corrections for anisotropic systems and Coulomb truncation schemes for low-dimensional materials.
Finally, the favorable $\bk$-scaling of the present approach makes its extension to metallic systems especially promising.
This will require handling fractional occupations and incorporating the plasmon contribution to the finite-size correction.
These developments will further broaden the scope and efficiency of the present low-scaling $GW$ framework for realistic large-scale materials applications.

\begin{acknowledgement}

We thank Prof. Matthias Scheffler for many valuable discussions on the methodological developments and results presented in this work, and M.-Y. Z. is especially grateful for his generous support during his stay at the NOMAD laboratory at the FHI.
We also thank Yu Cao, Huanjing Gong, and Bohan Jia for fruitful discussions. We acknowledge the funding support from the National Natural Science Foundation of China (Grants Nos.
12374067, 12134012, and 12188101) and the Strategic Priority Research Program of the Chinese Academy of Sciences 
(Grant No. XDB0500201).
This work was partially supported by the IOP-Humboldt Postdoctoral Fellowship in Physics of Institute of Physics, Chinese Academy of Sciences (Grant No. 202402).
This work was also funded by the National Key Research and Development Program of China (Grant Nos. 2022YFA1403800 and 2023YFA1507004) and the robotic AI-Scientist platform of the Chinese Academy of Sciences.

\end{acknowledgement}

\bibliography{main}

\begin{appendix}

\setcounter{table}{0}
\setcounter{figure}{0}
\renewcommand{\thetable}{A\arabic{table}}
\renewcommand{\thefigure}{A\arabic{figure}}

\section{Fourier transform in imaginary time/frequency domain}\label{app:ft}
In this work, we define the Fourier transform for functions of imaginary time/frequency as
\begin{equation}
\tilde{A}(\ii\omega) = \int_{-\infty}^{\infty} \dd{\tau}\, A(\ii\tau)\, \ee^{\ii \omega\tau},
\end{equation}
with the corresponding inverse transform
\begin{equation}
A(\ii \tau) = \int_{-\infty}^{\infty} \frac{\dd{\omega}}{2\pi}\, \tilde{A}(\ii \omega)\, \ee^{-\ii \omega\tau}.
\end{equation}
Within the zero-temperature formalism adopted here, imaginary-time quantities, such as the self-energy and Green's function, are not periodic in time.

For an even function $f$, the Fourier transform can be simplified to a cosine transform.
Its discretized form is written as
\begin{subequations}\label{eq:ct}
\begin{equation}\label{eq:ct-t-w}
\tilde{f}\left(\ii\omega_k\right) = \sum_{j=1}^N \gamma_{kj} \cos\left(\omega_k \tau_j\right) f\left(\ii \tau_j\right)
\end{equation}
\begin{equation}\label{eq:ct-w-t}
f\left(\ii\tau_j\right) = \sum_{k=1}^N \xi_{jk} \cos\left(\tau_j \omega_k\right) \tilde{f}\left(\ii \omega_k\right)
\end{equation}
\end{subequations}

Similarly, for an odd function $g$, the Fourier transform reduces to a sine transform, which is discretized as
\begin{subequations}\label{eq:st}
\begin{equation}\label{eq:st-t-w}
\tilde{g}\left(\ii\omega_k\right) = \ii\sum_{j=1}^N \lambda_{kj} \sin\left(\omega_k \tau_j\right) g\left(\ii \tau_j\right)
\end{equation}
\begin{equation}\label{eq:st-w-t}
g\left(\ii\tau_j\right) = -\ii\sum_{k=1}^N \zeta_{jk} \sin\left(\tau_j \omega_k\right) \tilde{g}\left(\ii \omega_k\right)
\end{equation}
\end{subequations}

\section{Lattice structures}\label{app:lattice}

This appendix collects the structural information for the materials considered in this work.
Table~\ref{tab:lattice} lists their space groups and lattice constants used in the calculations.

\begin{table}[!ht]
\centering
\caption{
 Crystal structures and lattice constants of the materials considered in this work.
  }%
\label{tab:lattice}
\begin{threeparttable}
\begin{tabular}{llr}
\toprule
Material & Space group & $a$ [\AA] \\
\midrule
AlP  & $F\bar{4}3m$ (\#216) &   5.4506    \\
AlAs & $F\bar{4}3m$ (\#216) &   5.6596    \\
BN   & $F\bar{4}3m$ (\#216) &   3.6148    \\
BP   & $F\bar{4}3m$ (\#216) &   4.5378    \\
C    & $Fd\bar{3}m$ (\#227) &   3.5668    \\
CaO  & $Fm\bar{3}m$ (\#225) &   4.7990    \\
CdS  & $F\bar{4}3m$ (\#216) &   5.8180    \\
GaN  & $F\bar{4}3m$ (\#216) &   4.5197    \\
GaP  & $F\bar{4}3m$ (\#216) &   5.4497    \\
GaAs & $F\bar{4}3m$ (\#216) &   5.6500    \\
GaSb & $F\bar{4}3m$ (\#216) &   6.0954    \\
InP  & $F\bar{4}3m$ (\#216) &   5.8696    \\
MgO  & $Fm\bar{3}m$ (\#225) &   4.2128    \\
MgS  & $Fm\bar{3}m$ (\#225) &   5.2027    \\
Si   & $Fd\bar{3}m$ (\#227) &   5.4164    \\
SiC  & $F\bar{4}3m$ (\#216) &   4.3578    \\
LiF  & $Fm\bar{3}m$ (\#225) &   4.0098    \\
LiCl & $Fm\bar{3}m$ (\#225) &   5.1295    \\
NaF  & $Fm\bar{3}m$ (\#225) &   4.6200    \\
NaCl & $Fm\bar{3}m$ (\#225) &   5.6397    \\
KF   & $Fm\bar{3}m$ (\#225) &   5.3470    \\
KCl  & $Fm\bar{3}m$ (\#225) &   6.2929    \\
ZnS  & $F\bar{4}3m$ (\#216) &   5.4188    \\
ZnO  & $F\bar{4}3m$ (\#216) &   4.5797    \\
\bottomrule
\end{tabular}
\end{threeparttable}
\end{table}

\end{appendix}

\end{document}